\documentclass[10pt]{article}

\usepackage{amsmath,amssymb}

\begin{document}
\setcounter{page}{0}
\thispagestyle{empty}
\begin{flushright}
GUTPA/05/01/01
\end{flushright}
\renewcommand{\thefootnote}{\fnsymbol{footnote}}
\begin{center}{\LARGE{{\bf Derivation of Poincare
Invariance from general quantum field theory}}}\end{center}
\vspace{1mm}
\begin{center}{\bf \large{C. D.
Froggatt}}
\end{center}
\renewcommand{\thefootnote}{\arabic{footnote}}
\begin{center}{{\it  Department of Physics and Astronomy,
University of Glasgow,  Glasgow G12 8QQ, Scotland}
}\end{center}
\vspace{1mm}
\begin{center}{\bf \large{H. B. Nielsen}}
\end{center}
\renewcommand{\thefootnote}{\arabic{footnote}}
\begin{center}{{\it  Niels Bohr Institute, Blegdamsvej 
17-21, DK 2100 Copenhagen, Denmark}
}\end{center}
\setcounter{footnote}{0}
\vspace{1mm}
\begin{center}
{\large \bf Abstract}
\end{center}
{\large
Starting from a very general quantum field theory we seek to
derive Poincar\'{e} invariance in the limit of low energy excitations.
We do {\em not}, of course, assume these symmetries at the outset,
but rather only a very general second quantised model. Many of the
degrees of freedom on which the fields depend turn out to
correspond to a higher dimension. We are not yet perfectly
successful. In particular, for the derivation of translational
invariance, we need to assume that some background parameters,
which {\em a priori} vary in space, can be interpreted as
gravitational fields in a future extension of our model. Assuming
translational invariance arises in this way, we essentially obtain
quantum electrodynamics in just $3+1$ dimensions from our model.
The only remaining flaw in the model is that the photon and the
various Weyl fermions turn out to have their own separate metric
tensors.
}

\newpage

\section{Introduction}
One of the most shocking results of Einstein's theory of
relativity, which again and again make people feel mystified, is
the relativity of time or of simultaneity. With our prejudice
about time, we have a natural tendency to believe in an absolute
time. In the present work we shall present a model which has an
absolute time without either Lorentz invariance or rotational
invariance at the outset. Nevertheless, under very general
assumptions, it essentially leads to the familiar Poincar\'{e}
invariant quantum electrodynamics.

This work is to be considered part of our long-standing ambitious
project of Random Dynamics \cite{book}, which attempts to
{\em derive} rather
than assume the laws of nature, preferably from an ``extremely
complicated fundamental world machinery''. That is really to say
that we assume that the fundamental laws of nature, or the
fundamental world machinery we could say, is so {\em enormously
complicated} that it would be hopeless to attempt to guess what it
is and write it down in detail. The best one could hope for might
be to give an idea of what type it is, but it would have so
enormously many unknown parameters that we would, in practice,
have to just assume that these parameters were random and hope
that this would not matter. In this sense, the Random Dynamics
philosophy consists of taking the fundamental world machinery to
be so complicated and so unspecified that, in reality, we have
assumed close to nothing. Then we hope that, in spite of assuming
that it almost does not matter what this world machinery is and
what its parameters are, we can get out {\em in some limits} the
laws of nature as we know them. We should consider the present
work as one little link in the long chain of deductions, in
principle leading from the random world machinery to the present
laws -- i.e.~the Standard Model of elementary particle physics.

\subsection{Central point of the article}
\label{central}

With our attempts to derive ``everything'' the article gets long
and complicated. The spirit of the article is to attempt to derive
properties that are normally simply postulated and taken for
granted. For example we shall spend some time discussing the
origin of the rudiments of translational invariance. So let us
here summarize the main points in our argument for Lorentz
invariance and for the number of spatial dimensions being
effectively three for the ``poor physicist'', who can only use
very low energies per particle compared to the fundamental
(Planck) mass scale. Such a physicist can safely Taylor expand to
lowest order in the momentum $\vec{p}_{phys}$ around that value
for which the available single particle states actually have their
energy approach zero in the boson case, or the Fermi surface in
the fermion case\footnote{Note that here boson and fermion purely
refer to the statistics of the particles. At this stage we are not
even assuming rotational invariance, let alone the spin-statistics
theorem.}. Let us, as shall anyway be done below, ``renormalise''
the momentum so that we have zero energy for zero momentum, by
shifting the momentum notation by an additive constant, and expand
around zero momentum. We shall generally argue that the equation
of motion, for the fields in momentum representation, takes the
form of the energies being given by the eigenvalues of an {\em
antisymmetric real} matrix, when we work with real fields --
corresponding to thinking of the single particle wave function as
being real, i.e. separated into real and imaginary parts in
momentum representation. Let us also here imagine, for the
illustration of our main point, that we have already managed to
argue for fermions that we effectively only have to consider four
real fields and for bosons only three. Let us even assume in the
fermion case that -- as we shall see -- the only relevant part of
the four by four matrix is the piece which commutes with the
matrix describing the single particle Hamiltonian with zero
momentum. Really we shall see in section \ref{Weylderivation} that
this zero momentum Hamiltonian functions as the imaginary element
$i$. It follows, in the fermion case, that instead of an
antisymmetric real four by four matrix we can work effectively
with an Hermitian two by two matrix.

That is to say we shall effectively argue for field equations of
the form
\begin{equation}
\dot{\xi_i}(\vec{p})= \sum_k A_{ik}\xi_k(\vec{p})
\end{equation}
where the $\xi_i$ refer to the fermion or boson fields, with the
index $i$ running {\em a priori} over all the many at first existing
Hermitian or real fields, but effectively taken to run only over 3
for bosons and 4 for fermions. Furthermore, in the fermion case,
we may even shift to having instead only two $i$ values but, to
make up for this, the fields become complex. The matrix elements
of the above mentioned antisymmetric real matrix are here denoted
$A_{ik}$ and they are functions of the momentum in a momentum
conserving theory. (Here momentum conservation is just put in for
pedagogical simplicity; in much of the present article we seek to
work without assuming momentum conservation.)

One of our main achievements is the prediction of the
dimensionality of space to be three -- with the one time dimension
in addition already assumed at the outset. This prediction arises
from the fact that, in the boson case, there are just three
linearly independent 3 x 3 real antisymmetric matrices and,
similarly in the fermion case, three linearly independent 2 x 2
Hermitian matrices in addition to the unit matrix.

For instance, by an appropriate basis (coordinate) choice in
momentum space, the Taylor expansion of the 3 x 3 real
antisymmetric matrix for small momenta can be expressed in the
form:
\begin{equation}
\label{eq2}
A = \left ( \begin{array}{c|c|c}0 & p_{3}& -p_{2}\\
-p_{3} & 0 & p_{1}\\ p_{2} & -p_{1} & 0\\
\end{array} \right)
\end{equation}
Although the boson fields considered here have three components
which, as already stressed, are real (or Hermitian when second
quantized), we have a trick one could say of getting {\em both
a magnetic and an electric field} out from them. We shall see that,
in the position or {\em $\vec{x}$-representation},
it corresponds to a
splitting into real and imaginary parts. This trick involves
letting the universe become an orbifold, or equivalently of
assuming the existence of a point around which the world is in a
parity symmetric {\em state}.

The crucial observation in our work is that, by this sort of most
general low momentum Taylor expansion, we actually happen to get
well-known {\em Lorentz invariant} equations of motion for the
fields, such as the Maxwell equations for the boson case and the
Weyl equation for the fermion case. In fact if we denote the three
components of the real boson field in momentum representation
as a three vector $\vec{\phi}(\vec{p})$, we may
interpret it in terms of momentum representation electric and
magnetic fields by
\begin{equation}
\vec{\phi} = \vec{\tilde{B}} +i\vec{\tilde{E}}.
\end{equation}
This equation may look a bit strange at first, when we stress so
much the reality of $\vec{\phi}$, but one should remember that the
Fourier transformed $\vec{\tilde{B}}$ and $\vec{\tilde{E}}$ are
{\em not} real even if the $\vec{x}$-representation electric and
magnetic fields are real. With the funny orbifold picture we argue
should appear in our model, we would rather get for example that
$\vec{\tilde{E}}$ is purely imaginary.

The present article is built around this derivation of the already
so well-known and empirically so successful equations of motion by
a Taylor expansion in momentum, starting from an exceedingly
general field theory. We first describe how we are led, with an
appropriate interpretation in terms of background gravitational
fields, to a kind of formal translational invariance. Then we give
a speculative argument for the number of relevant field components
at low energy. After deriving the free Maxwell and Weyl equations,
we allow for interactions. By imposing restrictions from the
required self-consistency of the equations of motion with
interactions included, we get very close to quantum
electrodynamics in a background gravitational field. However it
must be admitted there are still small flaws, the major one of
which is that the electron and photon get different metric
tensors. This may mean we have two gravity fields so far, but we
hope to progress further with this approach.

\subsection{Plan of the article}

In the following section \ref{general} we shall take as our
starting point a very general quantum field theory, so general
that we do {\em not} at first assume the usual physical principles
such as translational invariance and Lorentz
(including rotational) invariance. Indeed we do not really
even assume that there is a space; rather we assume that the
fields depend on a number of parameters, which are initially
taken to be greater than 3. One of our main results is precisely
to derive the existence of a space effectively having just
3 dimensions.
For simplicity and to be specific, we do though assume an absolute
time, which we hope is not so important in our model, and time
translational invariance. In section
\ref{singularity} we seek to argue that the two (and higher) point
functions in the very general quantum field theory must
effectively have the delta-function structure of just the same
type as in momentum conserving theories. However this does not yet
correspond to momentum conservation, but only to the absence of
potential terms in the integration kernels of the theory which do
not have this singularity structure. We give some speculative
hopes and an explanation of why we cannot expect, at least from
general arguments, to explain away the possibility of some
background field such as a background gravitational field.  In
section \ref{twopointkernels} we then write useful expressions for
the two point kernels with the suggested singularities and study
their symmetry and Hermiticity properties, especially formulated
in terms of commutator functions, anti-commutator functions and
single particle Hamiltonian functions extended to even be
considered as functions on the single particle phase space. Actually
an important point is that, due to our assumption that we should
work with real fields (or rather Hermitian fields in the second
quantized language), the phase space gets folded up as an orbifold
and we actually only get a ``half phase space'' -- or really
halved down once for each dimension. The most basic idea is to
Taylor expand the four functions, describing the two point kernels
and defined in section \ref{twopointkernels}, on phase space
in the very small range of momenta accessible, for practical
purposes, to a physicist working only with low energies compared
to the fundamental scale. In the next section \ref{normspec} we
choose the normalization of the fields, so as to simplify the
Hamiltonian in the case of bosons but the anti-commutator in the
case of fermions. It turns out that, in both cases, we end up with
having to seek eigenvalues for antisymmetric (and apart from an
over all factor $i$) real matrices.

In section \ref{zeropointeff} we present some effects for the very
general type of system we really consider: a system that is second
quantized with bosons and/or fermions and some degrees of freedom
that can adjust to lower the energy {\em under the influence
of the ``zero point energy''} of the second quantized degrees of
freedom. For fermions such  an effect is known -- by chemists --
under the name of Homolumo-gap. For the bosons we talk about the
``analogue of the Homolumo gap effect'' and the effect is that the
system will adjust itself so that very many boson frequencies (=
boson single particle energies) become exactly zero. The exactness
of the effect is connected with the fact that the zero point
energy has a ``numerical value singularity''.

Our derivations of Lorentz (and rotational) invariance in the free
boson and free fermion cases are put into sections
\ref{Maxwelleq} and \ref{Weylderivation} respectively.
For these derivations the zero-point
energy effects are quite important and, to be honest, we may even
use these effects in a bit too optimistic way.

In section \ref{chargeconservation} we call attention to the fact
that, in these type of models, one tends to get a charge
conservation symmetry as an accidental symmetry. This fact is
actually of importance in avoiding problems with the CPT-theorem,
due to level repulsion of the CPT-degenerate states which
otherwise would be expected. In a generic
theory -- as we work with -- one should not expect degeneracies to
occur without reason.

In section \ref{Interaction} we study the very general interaction
that might occur in our general field theory between the boson
fields, which we now interpret as the Maxwell fields, and the
fermion fields, which we interpret as the Weyl field for a
supposedly charged particle. We basically find that, in the
general Weyl equation, bosonic degrees of freedom appear, which
formally look like the gauge potential $A$. Then we study the
dominant commutator of this $A$-field with the Maxwell fields,
which we already obtained in the free approximation. A major
point, in arguing for the couplings we shall arrive at, is to
require consistency of the equations of motion with the
interaction included. In the last section \ref{conclusion} we
present the conclusion and some outlook.

\section{A very general quantum field theory}
\label{general}

We consider here a very generalized field theory model, in the
sense that {\em a priori} we do \underline{not} even have an
interpretation of the fields, $\phi_i(\vec{x}_{pre})$ say, as being
functions of a spatial point $\vec{x}_{pre}$. Instead, we rather think
of this ordered set of parameters $\vec{x}_{pre} =
(x^1_{pre},x_{pre}^2,...,x_{pre}^d)$ as
only ``some parameters on which the fields $\phi_i(\vec{x}_{pre})$
depend". For example, later they might turn out to be interpreted
as momentum rather than position and/or the true position or
momentum may turn out to be complicated functions of these
$\vec{x}_{pre}$-parameters and their conjugates $\vec{p}_{pre}$.
One might
think of these primitive parameters $\vec{x}_{pre}$ and their conjugate
single particle variables $\vec{p}_{pre}$ as kind of pre-position and
pre-momentum. In the following we shall in practice only consider
a small region of momentum space (i.e. the low energy region
accessible to experiment) and Taylor expand in momentum e.g.~terms
in the Hamiltonian, energies of particles, and the above-mentioned
complicated functions, which relate the primitive variables to the
single particle dynamical variables eventually identified with the
observed position and momentum.

In our generalized field theory, the field operators
$\phi_i(\vec{x}_{pre})$, which are strictly speaking operator-valued
distributions, act on the Hilbert space of the whole underlying
model. In addition to the considered fields, we imagine that the
full world machinery of our model contains other degrees of
freedom, which we shall refer to as the ``rest" degrees of
freedom. So the full Hilbert space is the Cartesian product of a
Hilbert space analogous to the Fock space in standard quantum
field theory and some further ``rest" Hilbert space.

In essentially the usual way, we shall consider continuous and
differentiable functions/distributions of the single particle
pre-position and pre-momentum variables
$\vec{x}_{pre}$ and $\vec{p}_{pre}$. By this we mean that the kernels,
(anti-) commutators etc.~appearing in the general theory
are assumed to be expressed in terms of genuine functions
and the most suggestive distributions, i.e.~the $\delta$-functions.
It should be stressed again here that, in the spirit of random
dynamics, we are considering a random and {\em a priori}
complicated model or world machinery. In particular, we do
\underline{not} put the usual translational or Lorentz symmetries
into the model. Rather our hope is to \underline{derive}
Poincar\'{e} invariance, as an effective or accidental symmetry
for a low energy observer.

For simplicity, we shall assume that the model has an absolute
time $t$ and a usual quantum mechanical development equation:
\begin{equation}
i\frac{\partial}{\partial t} |\psi> = H |\psi>
\label{schrodinger}
\end{equation}
where $|\psi>$ is a ket belonging to the Hilbert space. The very
general Hamiltonian is taken to be of the smooth form
\begin{equation}
\hspace{-18pt}
H = \sum_{n=0}^{\infty}\int K_{i_1\,i_2\,...\,i_n}(rest;
\vec{x}^{pre}_1,\vec{x}^{pre}_2,...,\vec{x}^{pre}_n)
\phi_{i_1}(\vec{x}^{pre}_1)
\phi_{i_2}(\vec{x}^{pre}_2)...\phi_{i_n}(\vec{x}^{pre}_n)
d^d\vec{x}^{pre}_1 d^d\vec{x}^{pre}_2 ... d^d\vec{x}^{pre}_n
\label{hamiltonian}
\end{equation}
as far as the fields are concerned. Here ``rest" in the kernel
$K$ indicates that $K$ would normally also be an operator in
the ``rest" Hilbert space.
Note that here and in the following, we use the subscript $pre$ and
the superscript $pre$ interchangeably.

\section{Derivation of translational invariance and locality,
at first only singularity structure} \label{singularity}

We now want to discuss how momentum conservation might arise
in such a general quantum field theory.
Since, as is well-known, momentum conserving kernels have the
singular form of a delta-function, say $\delta(\vec{p}_{pre} -
\vec{p'}_{pre})$ for the two-field case, we need to discuss the
singularity structure of the kernels $K_{i_1\,i_2\,...\,i_n}(rest;
\vec{x}^{pre}_1,\vec{x}^{pre}_2,...,\vec{x}^{pre}_n)$ in equation
(\ref{hamiltonian}) in order to at least argue for the right
singularity structure. {\em A priori} one could think of these
kernels as being smooth functions or smooth functions multiplied
by some products of $\delta$-functions. If the kernel had no
$\delta$-function with respect to $\vec{x}^{pre}_1$ say, then you
would only get a non-zero overlap with wave functions for a
limited range in the conjugate variable $\vec{p}^{pre}_1$.
However, it would then turn out that one of the terms in equation
(\ref{hamiltonian}) only depends on one of the fields $\phi_{i_1}$
for a small region in the single particle
($\vec{x}^{pre}_1,\vec{p}^{pre}_1$) phase space.

As this example of a smooth kernel shows, one can very easily risk to
get a kernel that projects away a huge number of states approximately.

One way of presenting the problem of avoiding the projection to
almost zero of a huge number of states is to imagine approaching
the continuous kernel from a lattice; say we replace the
continuous -- real -- variables or vectors $\vec{x}^{pre}_i$ by
discretized ones with some lattice distance $a$ between the points
on the coordinate axes which are actually included in the step
considered. Then, of course, the idea is to gradually make the
lattice distance smaller and smaller and at the end let $a
\rightarrow 0$. For example, we could think of this limit being
taken by successively diminishing the lattice constant by a factor
2, but really the details of how one makes the lattice more and
more dense should not matter for our argument.

\subsection{The no-go argument for smooth kernels without decoupling}

In order to show the decoupling for the great majority
of states when one has a smooth kernel, let us think of the
regularization of the operator produced by convolution with the
kernel $K_{i_1 i_2}(rest; \vec{x}_1^{pre},\vec{x}_2^{pre})$ as a
limit of a series of lattice replacement matrices. These
correspond to
replacing the wave functions defined on the vector space of
$\vec{x}_i^{pre}$-values by ``columns'' of numbers, with elements
being the wave function values at the sites in a lattice
introduced into this vector space; the kernel itself
is replaced by a matrix
with rows and columns marked by the lattice points. Denoting this
matrix associated with the kernel by ${\bf K}$,
the result of the
convolution
would be given as the limit of the lattice constant $a$ going to
zero of the matrix multiplication ${\bf K} {\bf \phi}$. The crux
of the argument now is that, as we go along diminishing $a$ and
thereby increasing the order of the matrix ${\bf K}$, we get
more and more eigenvectors and eigenvalues with
the added eigenvalues becoming smaller and smaller! We may estimate
the order of magnitude of the size of the added eigenvalues, by
estimating how the determinant varies as we add more and more
lattice points. Imagine that we simply add one more lattice point.
Then we add one row and one column to the matrix representing the
convolution ${\bf K}$. So its determinant is modified by a factor
given by what we may write in the added row, after having reduced
the numbers in this row by adding to it linear combinations of the
other (old) rows. Now, however, we can almost reproduce
the numbers in the new row corresponding to the new lattice point,
by Taylor expansion interpolations from the ``old'' lattice points,
because of the assumed smoothness of the kernel we
study. The interpolation estimate of a new row is a linear
combination of the other rows.
As the lattice gets more and more dense, the order to which we
can reproduce the Taylor series accurately gets
higher and higher, say up to a term involving the $p$th derivative.
Then
the size of the deviation of the true new row from its
interpolation estimate becomes of the order $a^p$ times the
typical size of the kernel values. This means that, as we add more
and more lattice points, the determinant gets suppressed by a
factor of the order $a^p$ for each extra
lattice point. Since the
determinant is the product of all the eigenvalues, this in turn
means that the added eigenvalues must be of the order $a^p$. That
in turn means that by far the majority of the eigenvalues of the
operator, corresponding to the convolution with the assumed
smooth kernel, are extremely small, since the factors
$a^p$ go to zero
very fast in the limit of the lattice becoming more and more
dense. We can also express this result by saying that by far the
majority of the wave functions, on which the smooth kernel acts by
convolution, are transformed into extremely small (essentially
vanishing) wave functions. Thus we can consider that most of the state
space is actually {\em decoupled} w.r.t. such a kernel.

It is really rather easy to understand what is going on,
by considering the various smooth functions
$K_{i_1\,i_2\,...\,i_n}(rest;
\vec{x}^{pre}_{10},\vec{x}^{pre}_{20},...,\vec{x}^{pre}_{(p-1)0},
\vec{x}^{pre}_p, \vec{x}^{pre}_{(p+1)0},...\vec{x}^{pre}_{n0})$,
obtained out of a kernel $K_{i_1\,i_2\,...i_n}$
(which could now depend on several variables) when
all the vector-variables are kept fixed, $\vec{x}_i =\vec{x}_{i0}$,
except for one $\vec{x}_p$. The resulting smooth functions
are approximately linearly dependent
upon each other due to interpolation. Therefore all the wave
functions $\phi_{i_p}(\vec{x}^{pre}_p)$
``orthogonal'' to the little subspace spanned by these
functions have essentially zero eigenvalues or, equivalently,
decouple. For by far the majority of wave functions we
can thus ignore the
smooth part of the kernels, without in practice causing much
disturbance.

\subsection{Next level of smoothness}

So now we have to weaken our smoothness assumption concerning the
kernels. Here we have to make a guess, based on aesthetics and general
experience, as to what type of functions one meets in physics next
to the obvious and usual smooth functions. We consider it fair to
say that, at the next level of smoothness, the types of
function which often occur physically are the Dirac delta-function
and its derivatives. In
order not to come into the problem with the smooth kernels, it is
necessary that there is a delta-function behaviour w.r.t.~any
one of the sets of variables on which the kernel $K$ depends. Let
us take the delta-function peak to be at the geometrical site for
which a certain function $\vec{f}(\vec{x}^{pre}_1,
...,\vec{x}^{pre}_n)$ is zero. Then the ``next type of function
in the series with different degrees of smoothness'' is obtained
as the delta-function
 $\delta (\vec{f}(\vec{x}^{pre}_1, ...,\vec{x}^{pre}_n))$
differentiated various numbers of times w.r.t. the various
variables and weighted with various genuinely smooth functions.
That is to say we can write the kernel
\begin{equation}
K=\sum_j F_j(\vec{x}^{pre}_1,...,\vec{x}^{pre}_n)
\delta^{(j)}(\vec{f}(\vec{x}^{pre}_1,...,\vec{x}^{pre}_n))
\end{equation}
where $j$ is an integer that runs from 0 to $\infty$, and the
symbol $(j)$ means the $j$th derivative.

One could easily get several terms of the form just described.
But if the ``poor physicist'' only sees the phase space locally,
he will only see one of the delta-functions and it will not be
important if there are more of them.

Now let us note that the derivatives on the delta-functions of
$\vec{f}$
could be produced by using derivatives w.r.t. any of the variables
$\vec{x}^{pre}_i$. In turn such derivatives are, apart from
a factor $i$, the operators corresponding to the conjugate variables
of the $\vec{x}^{pre}_i$'s.
We can thus replace the whole construction of derivatives {\em and}
the $F_j$ factors in front, by the action of an operator corresponding
to the quantum mechanical representation of some classical function
(defined on the phase space for all the variables $\vec{x}_i$ and their
conjugates). That is to say that we can write the general expression in
the form:
\begin{equation}
K= F(\vec{x}^{pre}_1,...,\vec{x}^{pre}_n,
\vec{p}^{pre}_1,...,\vec{p}^{pre}_n)
\delta(\vec{f}(\vec{x}^{pre}_1, ...,\vec{x}^{pre}_n)).
\end{equation}

At first we may see the seed for a conservation law in the
delta-function, since we could attempt to identify the $\vec{f}$
as the physical momentum. However, as we see, there are
derivatives of the delta-functions or, in the last equation, the
dependence on all the phase space variables, which would normally
spoil the translational invariance needed for momentum
conservation. Anyway we may perform the just mentioned step in
making it look like a momentum conserving kernel, by formally writing
$\vec{f}$ as a sum of ``physical momenta''
$\vec{p}_{phys,i}$ in the region accessible to the poor physicist
\begin{equation}
\vec{f}= \sum_i \vec{p}_{phys,i}.
\label{vecf}
\end{equation}
But, at first, we could not expect to get this simple form
(\ref{vecf}), if the fields depending on the different
$\vec{x}_i^{pre}$ and corresponding momenta $\vec{p}_i^{pre}$ are
defined on the same space of parameters $(\vec{x}_i^{pre},
\vec{p}_i^{pre})$. Rather there would be an additive constant
vector term in equation (\ref{vecf}). If the parameter spaces (or
phase spaces) are different, however, in the sense that the
different parameters $\vec{x}^{pre}_i$ are {\it a priori} of a
different nature and setting them equal would not be physically
meaningful, we can define the $\vec{p}_{phys,i}$ with such a
constant absorbed so as to really get equation (\ref{vecf}). The
crux of the matter is that the identification of the same point
for two different fields is only established by the use of the
singularity of the kernel itself. Now, in the ``poor physicist''
philosophy, the problematic constant vector will almost always be
huge on the ``poor physicist'' scale.  So the regions in phase
space, or momentum space, connected by the delta-function will be
so far apart that we can easily treat them as belonging
effectively to different spaces. According to this philosophy, we
should now really think of the typical situation as being that the
different field components $\phi_i$, $\phi_k$
are {\em a priori} defined on different small spots in phase or
momentum space, which are identified as relevant by the
delta-functions in the kernels. In this philosophy the field, with
a given index $i$, would not in practice come to multiply itself
(because if it did you would give it a new name) in say the
Hamiltonian (\ref{hamiltonian}). However once we begin to make
linear transformations on the fields with the same physical
momenta into each other, such diagonal terms can easily appear.

In this way we get that the delta-function looks momentum conserving,
but we also have the factor $F$ in the kernel which depends
on the conjugate variables to $\vec{p}_{phys,i}$. So the kernel
does not conserve momentum. It is rather locally of a
momentum conserving character,
but since it depends on position -- here we talk about the physical
position -- it is {\em not} translational invariant.
However we speculate
later that this $\vec{x}$ dependence could be ``dynamical''
(i.e.~depend on the ``rest'' variables) and somehow be adjusted
at low energy, or under some cosmological conditions, to
become effectively translational invariant after all.

\subsection{Commutators and anti-commutators}

In an analogous spirit we shall now argue for the singularity
structures for the commutators and anti-commutators.

Let us in fact consider the commutator
or anti-commutator (as the statistics for the fields will
require to be considered) of two fields $\phi_i(\vec{x}_{pre})$
and $\phi_j(\vec{x^{\prime}}_{pre})$:
\begin{equation}
[\phi_i(\vec{x}_{pre}), \phi_j(\vec{x^{\prime}}_{pre})]_{\pm} =
f_{ij}(\vec{x}_{pre}, \vec{x^{\prime}}_{pre})
+ \hbox{``field operators''}
\label{eq10}
\end{equation}
where we shall assume that the c-number term dominates
(this is like a free approximation would suggest).
From this commutator or anti-commutator we can
of course deduce, using the Fourier transformation
formulae
\begin{equation}
\phi_i(\vec{x}_{pre}) = \int \tilde{\phi}_i(\vec{p}_{pre})
\exp(i \vec{p}_{pre} \cdot \vec{x}_{pre})
\frac{d^d\vec{p}_{pre}}{(2\pi)^d},
\end{equation}
the commutator or anti-commutator of the
pre-$\vec{p}$
-space representation fields
\begin{equation}
[\phi_i(\vec{p}_{pre}), \phi_j(\vec{p^{\prime}}_{pre})]_{\pm} =
\tilde{f}_{ij}(\vec{p}_{pre}, \vec{p^{\prime}}_{pre}) +
\hbox{``field operators''}
\label{eq12}
\end{equation}
where
\begin{equation}
\tilde{f}_{ij}(\vec{p}_{pre}, \vec{p^{\prime}}_{pre})=
\int\exp{(-i\vec{p}_{pre} \cdot \vec{x}_{pre}
-i\vec{p^{\prime}}_{pre} \cdot \vec{x^{\prime}}_{pre})}
f_{ij}(\vec{x}_{pre},\vec{x^{\prime}}_{pre})
d^d\vec{x}_{pre}d^d\vec{x^{\prime}}_{pre}.
\end{equation}

In order to argue for the singularity structure of $f_{ij}$ or
$\tilde{f}_{ij}$, we require that weighted integrals over the
field $\phi_i(\vec{x}_{pre})$ taken to be a rapidly varying
function, $\int\phi_i(\vec{x}_{pre})
f_{ij}(\vec{x}_{pre},\vec{x^{\prime}}_{pre})d^d\vec{x}_{pre}$,
should not decouple in huge amounts, otherwise the field could be
said, in the above defined sense, to be ``effectively decoupled''.
Therefore we are {\em not} allowed to choose the functions
$f_{ij}$ and $\tilde{f}_{ij}$ in formulae (\ref{eq10}) and
(\ref{eq12}) to be so smooth as to have Taylor expansions in the
variables $\vec{p}_{pre}, \vec{p'}_{pre}, \vec{x}_{pre},
\vec{x'}_{pre}$. Rather, as we saw above, we need at least a
delta-function singularity as a function of (at least) one linear
combination (locally) of these variables. It is suggested that, as
a function of this combination variable, the behaviour should
correspond to a linear combination of the delta-function and its
derivatives with smooth coefficients. Such a form can in turn be
rewritten as an operator corresponding to a classical function in
phase space, which we may call $F_{ij}(\hbox{both }
\vec{x}'\hbox{s and } \vec{p}'\hbox{s})$. Indeed, by an
appropriate definition of physical momenta $\vec{p}_{phys}$ and
$\vec{p^{\prime}}_{phys}$, we may even write the delta-function as
$ \delta(\vec{p}_{phys} -\vec{p^{\prime}}_{phys})$. We thus manage
to write a commutator or anti-commutator as
\begin{equation}
[\tilde{\phi}_i(\vec{p}_{phys}),
\tilde{\phi}_j(\vec{p^{\prime}}_{phys})]_{\pm}
=F_{ij}(\vec{p}_{phys},\vec{x}_{phys}) \delta(\vec{p}_{phys} -
\vec{p^{\prime}}_{phys}) +...
\end{equation}
Again we can really only adjust the delta-function to be of
this form, provided we are allowed to redefine independently
the $\vec{p}_{phys}$ and  $\vec{p^{\prime}}_{phys}$. We can do
so if, in reality, the two fields $\tilde{\phi}_i(\vec{p}_{phys})$
and $\tilde{\phi}_j(\vec{p^{\prime}}_{phys})$ are defined
on different parameter spaces (or phase spaces), or if we
can take them to be effectively on different spaces.
This again means that we imagine the fields relevant for the
``poor physicist '' to be really
defined on small spots here and there in the fundamental
phase space.
But it should be stressed that the coefficient
function $F_{ij}$ is now a function of both momentum {\em and position}.
There is thus at this stage {\em no momentum conservation} yet.

It would also have been quite too much, if we could indeed have
derived momentum conservation properly just from the conditions
that there be no effective decoupling of a huge number of single
particle states. In fact the usual commutators or anti-commutators
modified by some, say, gravitational field -- which due to a
normalization rule for the fields has been moved from the
Hamiltonian into the commutator as is done in section
\ref{normspec} -- should have no such decoupling
problem. It should therefore not be possible to exclude background
fields in such a way. We only got here -- what is already not so
trivial -- the usual form with a delta-function for the
commutators, but translational invariance would require some
reason why possible fields that could modify them have adjusted
themselves to give a translational invariant background. We shall
see later that it is suggestive to speculate that these
coefficient functions $F_{ij}$ are dynamical and, thus, make the
question of their translational invariance much of the same type
of logic as to why our space-time is flat in general relativity.

\section{General field theory model for free theories}
\label{twopointkernels}

We shall formulate the model in terms of fields that
are essentially real or obey some Hermiticity conditions, which
mean that we can treat the bosonic and fermionic fields
$\phi(\vec{p}_{pre})$ and $\psi(\vec{p}_{pre})$ respectively
as Hermitian fields:
\begin{eqnarray}
\psi^{\dagger}(\vec{p}_{pre})& =& \psi(\vec{p}_{pre}),\\
\phi^{\dagger}(\vec{p}_{pre})& =& \phi(\vec{p}_{pre}).
\label{fieldHermiticity}
\end{eqnarray}
However the $\vec{p}_{pre}$ may not at the
end be the true momentum but rather what we could call pre-momentum.
In any case, one can always
split up a non-Hermitian field into its Hermitian and
anti-Hermitian parts. This is done since, in the spirit of the
random dynamics project, we do not want to assume any charge
conservation law from the outset. As we shall see in section
\ref{chargeconservation}, however, we have a mechanism that at least
would easily produce a charge conservation if we in any way should
get momentum conservation. The real reason we need
this Hermiticity
of the fields as a function of the pre-momentum is to get some
restrictions on the matrix of commutators in the boson case.

At the first stage in the development of our model, it is assumed
that we {\em only work to the free field approximation} and thus
the Hamiltonian is taken to be bilinear in the Hermitian fields
$\psi(\vec{p}_{pre})$ and $\phi(\vec{p}_{pre})$. Also possible terms
proportional to dynamical fields in the commutators or
anti-commutators will be ignored in the same approximation. In
conventional formulations one usually normalizes the fields to make
the commutators and anti-commutators simple and trivial. However,
since we shall here only follow
this conventional procedure for the fermion fields but not for the
bosons, it should be understood that to ignore
terms in the commutator (or anti-commutator) is also a free theory
approximation in our notation.

Let us first work in the pre-momentum representation, in which the
fields are Hermitian, and let us write down
the free approximation expressions for the Hamiltonian, the
anti-commutators and commutators for both fermions
and bosons.

Our Hamiltonians take the following forms:
\begin{equation}\label{FHam}
H_F =\frac{i}{2} \int d\vec{p}_{pre}d\vec{p'}_{pre} \:
\sum_{i,j}\psi_i(\vec{p}_{pre})\psi_j(\vec{p'}_{pre})
H_{ij}^{(F)}(\vec{p}_{pre},\vec{p'}_{pre})
\end{equation}
and
\begin{equation}\label{Ham}
H_B =\frac{1}{2} \int d\vec{p}_{pre}d\vec{p'}_{pre} \:
\sum_{i,j}\phi_i(\vec{p}_{pre})
\phi_j(\vec{p'}_{pre})H_{ij}^{(B)}(\vec{p}_{pre},\vec{p'}_{pre})
\end{equation}
for fermions and bosons respectively. Note we are
using the convention that, under the complex conjugation of
expressions like $H_F$, the order of fermionic quantities
is reversed; so we need the factor of $i$ in equation (\ref{FHam})
in order to make $H_{ij}^{(F)}$ real. The anti-commutators and
commutators take the forms:
\begin{equation}
\label{eqs18}
\{\psi_i(\vec{p}_{pre}),\psi_j(\vec{p'}_{pre})\}=
C_{ij}^{(F)}(\vec{p}_{pre},\vec{p'}_{pre})
\end{equation}
and
\begin{equation}
\label{eqs19}
[\phi_i(\vec{p}_{pre}), \phi_j(\vec{p'}_{pre})]
= iC_{ij}^{(B)}(\vec{p}_{pre},\vec{p'}_{pre})
\end{equation}
for fermions and bosons respectively.
According to our arguments above, the functions
$C^{(F)}_{ij}(\vec{p}_{pre},
\vec{p'}_{pre})$, $C^{(B)}_{ij}(\vec{p}_{pre},\vec{p'}_{pre})$,
$H^{(B)}_{ij}(\vec{p}_{pre},\vec{p'}_{pre})$  and
$ H^{(F)}_{ij}(\vec{p}_{pre},\vec{p'}_{pre})$ are singular
as delta-functions
or derivatives thereof so that, for example, we have the expansions
\begin{equation}
C^{(B)}_{ij}(\vec{p}_{pre},\vec{p'}_{pre}) =
\sum_L C^{(B)}_{Lij}(\vec{p}_{pre})\delta^{(L)}
(\vec{p}_{pre} - \vec{p'}_{pre}) =
C^{(B)}_{ij}(\vec{p}_{pre}, \vec{x}_{pre})
\delta(\vec{p}_{pre}- \vec{p'}_{pre})
\label{eq31}
\end{equation}
and
\begin{equation}
H^{(F)}_{ij}(\vec{p}_{pre},\vec{p'}_{pre}) =
\sum_L H^{(F)}_{Lij}(\vec{p}_{pre})\delta^{(L)}
(\vec{p}_{pre} - \vec{p'}_{pre}) =
H^{(F)}_{ij}(\vec{p}_{pre}, \vec{x}_{pre})
\delta(\vec{p}_{pre} - \vec{p'}_{pre}).
\label{eq32}
\end{equation}
Here the summation over the symbol $L$ is to be understood as the
summation over a set of non-negative integers, denoting the
numbers of differentiations on the delta-function to be performed
in the various components of the vectorial variable
$\vec{p}_{pre}$ or $\vec{p'}_{pre}$. In the last item of these two
expansions, the symbols $C^{(B)}_{ij}(\vec{p}_{pre},
\vec{x}_{pre})$ and $H^{(F)}_{ij}(\vec{p}_{pre}, \vec{x}_{pre})$
are now to be understood as quantum mechanical operators; so that
$\vec{x}_{pre}$ is really here interpreted as a differential
operator w.r.t. $\vec{p}_{pre}$. Strictly speaking they should
thus have had different symbols to $
C^{(B)}_{ij}(\vec{p}_{pre},\vec{p'}_{pre})$ and $
H^{(F)}_{ij}(\vec{p}_{pre},\vec{p'}_{pre})$, which are rather the
kernels corresponding to the operators
$C^{(B)}_{ij}(\vec{p}_{pre}, \vec{x}_{pre})$ and
$H^{(F)}_{ij}(\vec{p}_{pre}, \vec{x}_{pre})$. These operators
$C^B$, $C^F$, $H^B$ and $H^F$ are quantised versions of their
classical approximations, which are of course taken as smooth
functions over phase space.

It is obvious that we can -- for the $H$'s -- or  must
-- for the $C$'s -- choose these matrices so that they obey the
matrix symmetry relations:
\begin{eqnarray}
 H_{ij}^{(F)}(\vec{p}_{pre}, \vec{x}_{pre})&=&
-  H_{ji}^{(F)}(\vec{p}_{pre},\vec{x}_{pre}) \hbox{,}\quad\quad
\label{firstantisymmetry}\\
 H_{ij}^{(B)}(\vec{p}_{pre}, \vec{x}_{pre})&=&
H_{ji}^{(B)}(\vec{p}_{pre},\vec{x}_{pre}) \hbox{, } \\
 C_{ij}^{(F)}(\vec{p}_{pre}, \vec{x}_{pre})&=&
C_{ji}^{(F)}(\vec{p}_{pre},\vec{x}_{pre}) \quad\quad  \hbox{ and}\\
C_{ij}^{(B)}(\vec{p}_{pre}, \vec{x}_{pre})&=&
-  C_{ji}^{(B)}(\vec{p}_{pre},\vec{x}_{pre}).
\label{antisymmetry}
\end{eqnarray}
It is also obvious that, according to the equations (\ref{eqs18})
and (\ref{eqs19}), for each single pair of values ($i$, $j$) the
second quantized operators
$C_{ij}^{(B)}(\vec{p}_{pre},\vec{p'}_{pre})$ and
$C_{ij}^{(F)}(\vec{p}_{pre},\vec{p'}_{pre})$ must be Hermitian,
meaning real since they are c-numbers to leading order. In turn
this means that the single quantized operators
$C^{(B)}_{ij}(\vec{p}_{pre}, \vec{x}_{pre})$ and
$C^{(F)}_{ij}(\vec{p}_{pre}, \vec{x}_{pre})$ associated with them
by, for example, equation (\ref{eq31}) must be Hermitian. The
hermiticity properties of the analogous operators
$H^{(F)}_{ij}(\vec{p}_{pre}, \vec{x}_{pre})$ and
$H^{(B)}_{ij}(\vec{p}_{pre}, \vec{x}_{pre})$ given, for example by
equation (\ref{eq32}), are inherited from the expansion of the
Hamiltonians as in (\ref{Ham}). Since the second quantized fields
$\phi_i(\vec{p}_{pre})$ and $\psi_i(\vec{p}_{pre})$ are assumed to
be Hermitian (real in the first quantized sense), there would be
no reason to introduce anti-Hermitian terms which just cancel out
in the Hermitian Hamiltonian. We can therefore just take the
operators $H^{(F)}_{ij}(\vec{p}_{pre}, \vec{x}_{pre})$ and
$H^{(B)}_{ij}(\vec{p}_{pre}, \vec{x}_{pre})$ associated with the
matrices in (\ref{FHam}) and (\ref{Ham}) to be Hermitian (first
quantized wise) for each pair of indices ($i$,$j$).

Even in addition to the above relations, we have
the possibility of choosing the various functions ($H$'s and $C$'s)
to be invariant under the operation with the anti-linear first
quantized operator $PT_{pre}$, defined as simple complex
conjugation of the wave function in the $\vec{p}_{pre}$
representation. The wave functions, which we are effectively
having for the single particle description, are in fact real wave
functions and thus invariant under this operation. We can, so to
speak, just as well average every operator under this operation with
$PT_{pre}$ anyway. For operators composed from $\vec{p}_{pre}$ and
$\vec{x}_{pre}$, such as the $H$'s and $C$'s, this requirement
simply means symmetry under the sign shift of $\vec{x}_{pre}$,
i.e. $\vec{x}_{pre} \rightarrow - \vec{x}_{pre}$. We can also
compose this operation with the previous ones and obtain
equations, which could alternatively be seen as a set of equations
needed to ensure the Hermiticity of the total Hamiltonian:
 \begin{eqnarray}
 H_{ij}^{(F){\dagger}}(\vec{p}_{pre}, \vec{x}_{pre})&=&
H_{ij}^{(F)}(\vec{p}_{pre},-\vec{x}_{pre})
\label{firstHermiticity}
\hbox{,}\\
 H_{ij}^{(B){\dagger}}(\vec{p}_{pre}, \vec{x}_{pre})&=&
 H_{ij}^{(B)}(\vec{p}_{pre},-\vec{x}_{pre})
 \hbox{, } \\
 C_{ij}^{(F){\dagger}}(\vec{p}_{pre}, \vec{x}_{pre})&=&
 C_{ij}^{(F)}(\vec{p}_{pre},-\vec{x}_{pre})
\quad\quad  \hbox{ and}\\
C_{ij}^{(B){\dagger}}(\vec{p}_{pre}, \vec{x}_{pre})&=&
C_{ij}^{(B)}(\vec{p}_{pre},-\vec{x}_{pre}).
\label{Hermiticity}
\end{eqnarray}

Notice the minus signs on the $\vec{x}_{pre}$ arguments on the right
hand side of these equations. They come about because -- once the
symmetry or antisymmetry under the permutation of the indices
$i,j$ takes care of the anti-commutation in the $(F)$ case -- the
expressions like (\ref{eq31},\ref{eq32})
must be Hermitian, or real if we take them as c-number
functions of the two pre-momenta $(\vec{p}_{pre},\vec{p'}_{pre})$. So
simple derivatives, without any extra $i$ coefficients, are what is
needed in the terms corresponding, for example, to
$H^{(F)}_{Lij}(\vec{p}_{pre})$ and
$C^{(B)}_{Lij}(\vec{p}_{pre})$ in the sum over the different levels of
differentiation $L$. Thus, when we replace the derivatives
by the operator $\vec{x}_{pre}$, we insert in reality an extra $i$ for
which we must compensate by replacing
\begin{equation}
\vec{x}_{pre} \longrightarrow -\vec{x}_{pre}
\end{equation}
in the Hermiticity formulae.

We summarize the relations for the operators $H$'s and $C$'s
in the following way:

From the general considerations of the singularities we find that
the second quantized Hamiltonians take the form
\begin{eqnarray}
H_F& =&\frac{i}{2} \int d\vec{p}_{pre} \:
\sum_{i,j}\psi_i(\vec{p}_{pre})
H_{ij}^{(F)}(\vec{p}_{pre},\vec{x}_{pre})\psi_j(\vec{p}_{pre}),\\
H_B& =&\frac{1}{2} \int d\vec{p}_{pre} \:
\sum_{i,j}\psi_i(\vec{p}_{pre})
H_{ij}^{(B)}(\vec{p}_{pre},\vec{x}_{pre})\psi_j(\vec{p}_{pre}),
\end{eqnarray}
by combining equations (\ref{FHam},\ref{Ham}) with
(\ref{eq32}) and an analogous equation for $H_{ij}^{(B)}$.
Similarly the commutators of the second quantized fields are
obtained by inserting equation (\ref{eq31}) and its analogue, with
$B$ replaced by $F$, into equations (\ref{eqs18},\ref{eqs19}):
\begin{eqnarray}
\{\psi_i(\vec{p}_{pre}),\psi_j(\vec{p'}_{pre})\}& = &
C^{(F)}_{ij}(\vec{p}_{pre}, \vec{x}_{pre})
\delta(\vec{p}_{pre} - \vec{p'}_{pre})\\
\;  [ \phi_i(\vec{p}_{pre}),\phi_j(\vec{p'}_{pre})]& = &
iC^{(B)}_{ij}(\vec{p}_{pre}, \vec{x}_{pre})
\delta(\vec{p}_{pre} - \vec{p'}_{pre}).
\end{eqnarray}
Then the operators acting on the first quantized Hilbert
space\footnote{Here the Hilbert space is of the slightly unusual
type of being built on real wave functions only in the
$\vec{p}$-representation.} can be thought upon classically as
functions on the phase space (as discussed above, they are taken
to be smooth functions of the operators $\vec{p}_{pre}$ and the
conjugate differential operators $\vec{x}_{pre} =
i\frac{\partial}{\partial \vec{p}_{pre}}$) obeying the following
symmetry properties:
\begin{enumerate}
\item
All the matrix elements are real over the whole of phase space.
\item
The matrices $C^{(B)}$ and $H^{(F)}$ are antisymmetric, while the two
other matrices $C^{(F)}$ and $H^{(B)}$
are symmetric all over phase space.
\item
All of the four matrices take as functions over phase space the
{\em same} values for $(\vec{p}_{pre}, \vec{x}_{pre})$ and
for $(\vec{p}_{pre}, -\vec{x}_{pre})$, i.e. we have the relations
\begin{eqnarray}
H^{(B)}_{ij}(\vec{p}_{pre},\vec{x}_{pre}) =
H^{(B)}_{ij}(\vec{p}_{pre},-\vec{x}_{pre}); \quad &
H^{(F)}_{ij}(\vec{p}_{pre},\vec{x}_{pre}) =
H^{(F)}_{ij}(\vec{p}_{pre},-\vec{x}_{pre});\qquad &  \\
C^{(B)}_{ij}(\vec{p}_{pre},\vec{x}_{pre}) =
C^{(B)}_{ij}(\vec{p}_{pre},-\vec{x}_{pre}); \quad &
C^{(F)}_{ij}(\vec{p}_{pre},\vec{x}_{pre}) =
C^{(F)}_{ij}(\vec{p}_{pre},-\vec{x}_{pre}). \qquad &
\end{eqnarray}
\end{enumerate}

\subsection{Canonical Transformations}

We do not want to assume -- and perhaps are forced not to assume
if we want to get momentum conservation -- that the pre-momentum
$\vec{p}_{pre}$ is necessarily to be interpreted as the true
physical momentum. So we should like to allow for the possibility
that the concepts of momentum and position have been canonically
transformed relative to the pre-momentum and pre-position (for the
single particles). We should therefore like to contemplate what
happens to the (anti-)symmetry relations
(\ref{firstantisymmetry}-\ref{antisymmetry}) and the Hermiticity
relations (\ref{firstHermiticity}-\ref{Hermiticity}) under
canonical transformations  of the $(\vec{p}_{pre},\vec{x}_{pre})$,
corresponding to unitary transformations in the single particle
space. In fact we want to deduce that the number of restrictions
is not changed, although their form gets more complicated.

Thinking of a classical approximation to the four
functions $H^{(F)}_{ij},\ H^{(B)}_{ij},\ C^{(F)}_{ij},
\ C^{(B)}_{ij}$, it should be obvious that, under a canonical
transformation to new variables $(\vec{p}, \vec{x})$
in place of the ``fundamental'' ones
$(\vec{p}_{pre}, \vec{x}_{pre})$,
we transform the equations (\ref{firstantisymmetry}-\ref{Hermiticity})
into the same ones just expressed in terms of the new variables. So
equations (\ref{firstantisymmetry}-\ref{antisymmetry})
come to look quite the same, except that now
the variables are the new ones  $(\vec{p}, \vec{x})$
instead of the old ones. In order to formulate the equations
(\ref{firstHermiticity}-\ref{Hermiticity}) in terms of the physical
coordinates, we need to give a
name to the point into which the canonical transformation brings
the reflected $(\vec{p}_{pre}, -\vec{x}_{pre})$ point of
 phase space. Let us
call it $r(\vec{p}, \vec{x}) $, then equations
(\ref{firstHermiticity}-\ref{Hermiticity}) take the form
\begin{eqnarray}
 H_{ij}^{(F){\dagger}}(\vec{p}, \vec{x})&=&
H_{ij}^{(F)}(r(\vec{p},\vec{x}))
\hbox{,}\quad\quad
\label{firstgeneralHermiticity}\\
 H_{ij}^{(B){\dagger}}(\vec{p}, \vec{x})&=&
H_{ij}^{(B)}(r(\vec{p},\vec{x}))
 \hbox{, } \\
 C_{ij}^{(F){\dagger}}(\vec{p}, \vec{x})&=&
C_{ij}^{(F)}(r(\vec{p},\vec{x}))
\quad\quad \hbox{ and}\\
C_{ij}^{(B){\dagger}}(\vec{p}, \vec{x})&=&
  C_{ij}^{(B)}(r(\vec{p} ,\vec{x})).
\label{generalHermiticity}
\end{eqnarray}
Under this unitary/canonical transformation we, of course, also
transform the wave function for the single particle or,
equivalently, the fields into new fields. Thus we can now decide,
for example, to use the physical momentum representation in which
the ``physical momentum '' $\vec{p}$, defined below in subsection
\ref{ss44a49}, is diagonal. The fields in this representation are
naturally called $\phi_i(\vec{p}) $ and $ \psi_i(\vec{p})$. The
symmetry relations (\ref{firstantisymmetry}-\ref{antisymmetry})
will also be valid when we use the ``physical'' momenta and
positions instead of the $
(\vec{p}_{pre},\vec{x}_{pre})$-variables.

In the following subsection \ref{ss44a49} we shall discuss the way
we are supposed to select the ``physical'' variables and thus
which canonical transformation, or rather unitary transformation,
should be performed, in order to go into the variables to be
identified at the end with what we conceive to be the true momenta
and positions.

\subsection{The choice of what is physical position and momentum}
\label{ss44a49}

We shall study below the spectrum of frequency
eigenstates for the just discussed functions $H$ and $C$,
which formally look like functions defined on the phase space the
coordinates of which are the pre-momenta and pre-positions.
Quite abstractly and generally however, we may
consider a spectrum of frequency eigenvalues $\omega$ to vary
continuously over the phase space. Now the ``poor physicist'' has
only very small energies at his disposal compared to the
fundamental (presumably Planck) energy unit. So he only has access
to the part of the spectrum which is very close to the
Fermi-surface for fermions and close to zero frequency for the
bosons. As one goes along in the phase space, the spectrum
typically varies on a scale of the fundamental frequency.
Therefore, in most of phase space, it will be tremendously faraway
from the Fermi-surface or from zero from the point of view of the
``poor physicist''. (We shall below argue that some boson
frequencies remain exactly zero, but let us here for pedagogical
reasons ignore these eigenvalues in the spectrum, which are anyway
assumed not to be truly observed.) The eigenfrequencies which
reach the Fermi-surface or the zero frequency surface do so on a
sub-manifold of the phase space -- which remarkably enough we
shall show below to be of co-dimension 3 - and grow up along this
sub-manifold linearly with momentum (the dominant Taylor expansion
terms).
Only a narrow band in phase space around this co-dimension 3 surface
is relevant for the poor physicist.

Now the physical position and momentum, which henceforth will be
denoted by $\vec{x}$ and $\vec{p}$, are to be identified with
linear combinations of pre-position and pre-momentum, such that
the physical position varies along the bands whereas the physical
momentum varies across the bands. To be more precise, we shall see
below that in 3 directions the single particle frequencies
increase linearly with the distance from a co-dimension 3
sub-manifold. Along this co-dimension 3 sub-manifold we have an
extra $\omega = 0$ eigenvalue in the boson case and two
eigenstates at just the Fermi-surface in the fermion case. It is
these three directions along which the relevant $\omega$'s
increase linearly that are to be identified with the physical
momenta. We shall set the zero for the physical momentum to be
just at the above-mentioned co-dimension 3 surface. The physical
position must then be defined to be along this co-dimension 3
surface. In general this co-dimension 3 surface will have more
than 3 dimensions and, thus, the choice of the physical position
space {\em a priori} now becomes arbitrary. However we can choose
the genuine position directions to be determined from the physical
momenta, by using them in a Poisson bracket to generate
translations.

The physics that is relevant for the ``poor physicist'' is
only the behaviour of the accessible part of the spectrum
in the very narrow bands mentioned. We make the very general
assumption that, in these bands, all the relevant functions $H$
and $C$ are Taylor expandable with high order terms being discarded.

Really the general procedure in our attempts to
derive Lorentz invariance etc.~is to use Taylor expansions,
especially for the four functions we have written about here.
For example we may use the Taylor expansion
\begin{equation}
H^{(F)}_{ij}( \vec{p}, \vec{x}) \approx
H^{(F)}_{ij}( \vec{0}, \vec{x})+ \vec{p} \cdot
\frac{\partial}{\partial \vec{p}} H^{(F)}_{ij}(
\vec{p}, \vec{x})|_{\vec{p}=\vec{0}} +...
\end{equation}
because the physical $\vec{p}$, with which
the ``poor physicist'' can in practice work, is a very
small quantity compared to the fundamental scale.

When the four operators we study here have to obey a reality
condition like (\ref{firstHermiticity}-\ref{Hermiticity}), the
number of independent, say, first derivatives
$\frac{\partial}{\partial \vec{p}} H^{(F)}_{ij}( \vec{p},
\vec{x})|_{\vec{p}=\vec{0}}$ gets lowered. Really the number of
restrictions and thereby also the number of independent matrices
in the expansion should depend in a smooth continuous way on the
unitary transformation. Therefore, as one of course always expects
an integer number of restrictions, this number must be independent
of this deformation. In fact we may differentiate the equations
(\ref{firstantisymmetry}-\ref{antisymmetry}) with respect to
physical variables, since we remember that they are true also with
the physical position and momentum, and we may differentiate the
Hermiticity conditions
(\ref{firstgeneralHermiticity}-\ref{generalHermiticity}) to get
equations like
\begin{eqnarray}
\frac{\partial}{\partial \vec{p}}
C^{(F)}_{ij}&=
&\frac{\partial}{\partial \vec{p}} C^{(F)}_{ji},\\
\frac{\partial}{\partial \vec{p}}
C^{(F){\dagger}}_{ij}(\vec{p},\vec{x})&=&
\frac{\partial}{\partial \vec{p}}
C^{(F)}_{ij}(r(\vec{p},\vec{x})).
\end{eqnarray}
There are quite analogous equations for the other three
functions we consider.

\section{Normalization of the fields $\psi_i$ and $\phi_i$
by fixing the symmetric matrices to Kronecker deltas, equations of
motion and spectra} \label{normspec}

It should be borne in mind that {\it a priori} the fields
are arbitrarily normalized and that we may use the Hamiltonians to
define the normalization of the fields, if we so choose. In fact
an important ingredient in the formulation of the present work is
to assume that a linear transformation has been made on the
various field components $\phi_i(\vec{p})$, i.e.~a transformation
on the component index $i$, such that the symmetric coefficient
functions $H_{ij}^{(B)}(\vec{p},\vec{x})$ become equal to the
unit matrix:
\begin{equation}
H_{ij}^{(B)}(\vec{p},\vec{x}) = \delta_{ij}
\label{eq49}
\end{equation}
for the very small momenta accessible to the poor physicist.
Thereby, of course, the commutation relations
among these components $\phi_i(\vec{p})$ are modified and
we cannot simultaneously arrange for them to be trivial.
So for the bosons
we choose a notation in which the non-trivial behaviour of
the equations of motion, as a function of  the momentum
$\vec{p}$, is put into
the commutator expression
\begin{equation}
[\phi_i(\vec{p}), \phi_j(\vec{p'})]=
iC^{(B)}_{ij}(\vec{p}, \vec{x})
\delta(\vec{p} - \vec{p'})
\label{commutator}
\end{equation}
It follows that the information which we would, at first,
imagine should be contained in the Hamiltonian is, in fact,
now contained in the antisymmetric matrices
$C^{(B)}_{ij}(\vec{p}, \vec{x})$.

This means that we normalize the fields so as to make the
Hamiltonian become

\begin{equation}
H_B = \frac{1}{2}\int \sum_i \phi_i(\vec{p}) \phi_i(\vec{p})
d\vec{p}.
\end{equation}

In principle we could also have used the same normalization
procedure for the fermions. However, in both cases, we rather
choose the symmetric functions among our four $C_{ij}$ and
$H_{ij}$ functions to be normalized to the Kronecker deltas
in Eq.~ (\ref{eq49}). So, for the fermions, we
just take $C^{(F)}_{ij} =\delta_{ij}$.
This means that for the fermions,
we shall keep to the more usual formulation.
In other words, we normalize the anti-commutator to be the
unit matrix and let the more
non-trivial dependence on $\vec{p}$ sit in the Hamiltonian
coefficient functions $H_{ij}^{(F)}(\vec{p}, \vec{x})$.
That is to say that
we have the usual equal time anti-commutation relations:
\begin{equation}
\{ \psi_i(\vec{p}), \psi_j(\vec{p'})\} =
\delta_{ij} \delta(\vec{p} - \vec{p'}).
\end{equation}

The component indices $i$, $j$ enumerate the very general discrete
degrees of freedom in the model. These degrees of freedom might, at the
end, be identified with Hermitian and anti-Hermitian components, spin
components, variables versus conjugate momenta or even totally
different types of particle species, such as flavours etc.
It is important to realize that this model
is so general that it has, in that sense, almost no assumptions built
into it---except for our free approximation, the above-mentioned
rudimentary momentum conservation and some general features of
second quantized models. It follows from the rudimentary momentum
conservation in our model that the
(anti-)commutation relations have a $\delta(\vec{p} - \vec{p'})$ delta
function factor in them.

\subsection{Equations of motion for the general fields}

We can easily write down the equations of motion for the
field components
in our general quantum field theory, both in the fermionic case:
\begin{equation}
\dot{\psi}_i(\vec{p}) = i[H_F, \psi_i(\vec{p})] =
 - \sum_k\psi_k(\vec{p})H^{(F)}_{ki}(\vec{p}, \vec{x})
\end{equation}
and in the bosonic case:
\begin{equation}
\label{eq8} \dot{\phi}_i(\vec{p}) = i[H_B, \phi_i(\vec{p})]
= - \sum_k\phi_k(\vec{p})C^{(B)}_{ki}(\vec{p}, \vec{x}).
\end{equation}

We see that both the bosonic and the fermionic equations of motion
are of the form
\begin{equation}
\dot{\xi_i}(\vec{p})= \sum_k A_{ik}\xi_k(\vec{p})
\end{equation}
Here $A_{ij}(\vec{p}, \vec{x})$ is a real antisymmetric matrix and
we have introduced the neutral name $\xi$ for both the boson field
$\phi$ and the fermion field $\psi$.

\subsection{Spectrum of an antisymmetric matrix
$A_{ij}(\vec{p}, \vec{x})$}

An antisymmetric matrix $A_{ij}(\vec{p}, \vec{x})$ with real
matrix elements is anti-Hermitian and thus has purely
imaginary eigenvalues. However, if
we look for a time dependence ansatz of the form
\begin{equation}
\xi_i(\vec{p},t)= a_i(\vec{p}) \exp(-i\omega t),
\end{equation}
the eigenvalue equation for the frequency $\omega$ becomes
\begin{equation}
\omega a_i= \sum_j iA_{ij}(\vec{p}, \vec{x})a_j.
\end{equation}
Now the matrix $iA_{ij}(\vec{p}, \vec{x})$ is Hermitian and the
eigenvalues $\omega$ are therefore real.

It is easy to see, that if $\omega$ is an eigenvalue, then so also is
$-\omega$. In fact we could imagine calculating the eigenvalues by
solving the equation
\begin{equation}
\det{(iA - \omega)} =0.
\end{equation}
We then remark that transposition of the matrix $(iA-\omega)$ under the
determinant sign will not change the value of the determinant, but
corresponds to changing the sign of $\omega$ because of the
antisymmetry of the matrix $iA$.
So non-vanishing eigenvalues occur in pairs.

Now the main point of interest for our study is how the second
quantized model looks close to its
ground state.  The neighbourhood of this ground state is supposed to
be the only regime which we humans can study in our ``low energy''
experiments, with small momenta compared to the fundamental (say
Planck) mass scale. With respect to the ground state of such a
second quantized world machinery, it is well-known that there is
a difference between the fermionic and the bosonic case. In the
fermionic case, you can at most have one fermion in
each state and must fill the states up to some specific value of the
single particle energy---which is really $\omega$.
However, in the bosonic
case, one can put a large number of bosons into the same orbit/single
particle state, if that should pay energetically.

\subsection{The vacuum}
If we allow for the existence of a chemical potential, which
essentially corresponds to the conservation of the number of
fermions, we shall typically get the ground state to have fermions
filling the single particle states up to some special value of the
energy called the Fermi-energy $\omega_{FS}$ ($FS$ standing for
``Fermi-surface''). For bosons, on the other hand, we will always
have zero particles in all the orbits, except perhaps in the zero
energy ground state; it will namely never pay energetically to put
any bosons into positive energy orbits.

\subsection{The lowest excitations}

So for the investigation of the lowest excitations, i.e.~those that a
``poor physicist'' could afford to work with, we should look for
the excitations very near to the Fermi-surface in the fermionic
case. In other
words, we should put fermions into the orbits with energies very little
above the Fermi-energy,
or make holes in the Fermi-sea at values of the orbit-energies very
little below the Fermi-energy. Thus, for excitations accessible to
the ``poor physicist'', it is only necessary to study the behaviour
of the spectrum for the bosons having a value of $\omega$ near to zero,
and for the fermions having a value of $\omega$ near the Fermi-energy
$\omega_{FS}$.

\section{Jahn-Teller effect and analogue for bosons}
\label{zeropointeff}

The Jahn-Teller effect \cite{Teller} refers to a very general feature
of systems of fermions, which possess some degrees of freedom that
can adjust themselves so as to lower the energy as much as
possible. The effect is so general that it should be useful for
almost all systems of fermions, because even if they did not have
any extra degrees of freedom to adjust there would, in the Hartree
approximation, be the possibility that the fermions could
effectively adjust themselves. The name Homolumo gap was
introduced for this effect in chemistry and stands for the
gap between `` the
highest occupied'' HO ``molecular orbit'' MO and the ``lowest
unoccupied'' LU ``molecular orbit'' MO.  The point is simply that
if the filled (occupied) orbits (single particle states) are
lowered the whole energy is lowered, while it does not help to
lower the empty orbits. It therefore pays energetically to make
the occupied orbits go down in energy and separate from the unfilled
ones; thus a gap may appear or rather {\em there will be a general
tendency to get a low level density near the Fermi-surface}. This
effect can easily be so strong that it causes a symmetry to break
\cite{Teller}; symmetry breaking occurs if some levels, which are
degenerate due to the symmetry, are only partially filled so that
the Fermi-surface just cuts a set of degenerate states/orbits. It
is also the Homolumo-gap effect which causes the deformation of
transitional nuclei, which are far from closed shell
configurations. We want to appeal to this Homolumo gap
effect, in section \ref{Weylderivation}, as a justification for
the assumption that the Fermi-surface gets close to those places
on the energy axis where the level density is minimal.

However we first want to discuss a similar effect, where the degrees
of freedom of a system of bosons adjust themselves to lower the
total energy. As for the fermion systems just discussed, this
lowering of the total energy is due to the adjustment of a sum
over single particle energies---the minimisation of the zero-point
energy of the bosonic system. We consider the effect of this
minimisation to be the analogue for bosons of the Jahn-Teller
effect.

\subsection{The analogue for bosons}

Let us imagine that there are many degrees of freedom of the whole
world machinery that could adjust themselves (we called them the
``rest'' degrees of freedom in section \ref{general}) to minimize
the energy of the system and could also influence the matrix
$A_{ij}(\vec{p}, \vec{x})$. Then one could, for instance, ask how
it would be energetically profitable to adjust the eigenvalues, in
order to minimize the zero-point energy of the whole (second
quantized) system. This zero-point energy is formally given by the
integral over all (the more than three dimensional) momentum
space; let us just denote this integration measure by $d\vec{p}$,
so that:
\begin{equation}
\label{zpe}
E_{{\it zero-point}} = \int  d\vec{p}
\sum_{{\it eigenvalue\,\, pair   \,\, k}}
|\omega_k(\vec{p})|/2.
\end{equation}
Provided some adjustment took place in minimizing this quantity,
there would {\em a priori} be an argument in favour of having
several zero eigenvalues, since they would contribute the least to
this zero-point energy $E_{{\it zero-point}}$. At first sight,
this argument is not very strong, since it just favours making the
eigenvalues small and not necessarily making any one of them
exactly zero. However, we underlined an important point in favour
of the occurrence of exactly zero eigenvalues, by putting the
numerical sign explicitly into the integrand
$|\omega_k(\vec{p})|/2$ in the expression (\ref{zpe}) for the
zero-point energy. The important point is that the numerical value
function is not an ordinary analytic function, but rather has a
kink at $\omega_k(\vec{p})=0$. This means that, if other
contributions to the energy of the whole system are
smooth/analytic, it could happen that the energy is lowered when
$\omega_k(\vec{p})$ is lowered numerically for both signs of
$\omega_k(\vec{p})$; here we consider $\omega_k(\vec{p})$ to be a
smooth function of the adjusting parameters of the whole world
machinery (the ``rest'' degrees of freedom). For a
normal analytic energy function this phenomenon could of course
never occur, except if the derivative just happened (is fine-tuned
one could say) to be equal to zero at $\omega_k(\vec{p}) =0$.
But with a contribution that has the numerical value singularity
behaviour it is possible to occur with a finite probability
(i.e.~without fine-tuning), because it is sufficient that the
derivative of the contribution to the total energy from other
terms
is numerically lower than the derivative of the zero-point term
discussed. Then, namely, the latter will dominate the sign of the
slope and the minimum will occur {\em exactly} for zero
$\omega_k(\vec{p})$.

In this way, we claim to justify our assumption in the following
section that the matrix
$A_{ij}(\vec{p}, \vec{x})$ will have several,
say $n$ exactly zero eigenvalues
and thus a far from maximal rank; the rank being at least piecewise
constant over momentum space. We shall therefore now study
antisymmetric matrices with this property in general and look for
their lowest energy excitations.

\section{Derivation of Lorentz invariance and the free
Maxwell equations}
\label{Maxwelleq}

It can happen that, for special values of the ``momentum parameters'',
a pair of eigenvalues of the antisymmetric matrix
$A_{ij}(\vec{p}, \vec{x})$---consisting of eigenvalues
of opposite sign of
course---approach the group of exactly zero eigenvalues.
It is this situation which we believe to
be the one of relevance for the low energy excitations.
We shall therefore concentrate our interest on a small region in the
momentum parameter space, around a point $\vec{p}_0$ where the two
levels with the numerically smallest non-zero eigenvalues merge
together with $n$ levels having zero eigenvalue. Following
the notation for physical momentum introduced in section
(\ref{ss44a49}), we set $\vec{p}_0=0$.
Using the well-known
fact that, in quantum mechanics, perturbation corrections from
faraway levels have very little influence on the perturbation of a
certain level, we can ignore all the levels except the $n$ zero
eigenvalues and this lowest non-zero pair.

Hence we end up with an effective $(n+2) \times (n+2)$ matrix
$A_{ij}(\vec{p}, \vec{x})$ obeying the constraint of being of rank
two (at most). Now we imagine that we linearize the momentum
dependence of $A_{ij}(\vec{p}, \vec{x})$ on $\vec{p}$ around a
point in momentum space, $\vec{p}_{0}=0$, where the pair of
eigenvalues approaching zero actually reach zero, so that the
matrix is totally zero, $A_{ij}(\vec{p}_0, \vec{x})=0$, at the
starting point for the Taylor expansion. That is to say that,
corresponding to different basis vectors in momentum space, we get
contributions to the matrix $A_{ij}(\vec{p}, \vec{x})$ linear in
the momentum difference $\vec{p}-\vec{p}_0 =\vec{p}$. Now any
non-zero antisymmetric matrix is necessarily of rank at least 2.
So the contribution from the first chosen basis vector in momentum
space will already give a matrix $A_{ij}$ of rank 2 and
contributions from other momentum components should not increase
the rank beyond this. A single basis vector for a set of linearly
parameterised antisymmetric real matrices can be transformed to
just having elements (1,2) and (2,1) nonzero and the rest zero. In
order to avoid a further increase in the rank of the matrix by
adding other linear contributions, these further contributions
must clearly not contribute anything to matrix elements having
both column and row index different from $1$ and $2$. However this
is not sufficient to guarantee that the rank remains equal to 2.
This is easily seen, because we can construct $4 \times 4$
antisymmetric matrices, which are of the form of having 0's on
all places (i,j) with both i and j different from 1 and 2 and
have non-zero determinant.

So let us consider 4 by 4 sub-determinants of the matrix $A_{ij}$
already argued to be of the form
\begin{equation}
\label{mform}
\left ( \begin{array}{c|c|c|c|c}0 & A_{12} & A_{13} &\cdots & A_{1n}\\
-A_{12} & 0 & A_{23}& \cdots & A_{2n}\\ -A_{13} & -A_{23} & 0&
\cdots &0\\
: & : & 0 & \cdots &0\\
-A_{1n} & -A_{2n} & 0 & \cdots & 0 \end{array} \right).
\end{equation}
Especially let us consider a four by four sub-determinant along the
diagonal involving columns and rows 1 and 2. The determinant is
for instance
\begin{equation}
\label{sub4}
\det
\left ( \begin{array}{c|c|c|c}0 & A_{12} & A_{13} & A_{15}\\
-A_{12} & 0 & A_{23}& A_{25}\\ -A_{13} & -A_{23} & 0&0\\
-A_{15} & -A_{25} & 0 & 0 \end{array} \right ) =-\left( \det \left (
\begin{array}{c|c} A_{13} & A_{15}\\
 A_{23}& A_{25} \end{array} \right ) \right )^2.
\end{equation}
In order that the matrix $A_{ij}$ be of rank 2, this determinant
must vanish and so we require that the 2 by 2 sub-matrix
\begin{equation}
\left (
\begin{array}{c|c} A_{13} & A_{15}\\
 A_{23}& A_{25}\\ \end{array} \right)
\end{equation}
must be degenerate, i.e.~of rank 1 only. This means that the
two columns in it are proportional, one to the other. By
considering successively several such selected four by four
sub-matrices, we can easily deduce that all the two columns
\begin{equation}
\left ( \begin{array}{c} A_{13}\\ A_{23}  \end{array} \right),
\left ( \begin{array}{c} A_{14}\\ A_{24}  \end{array} \right), \cdots
\left ( \begin{array}{c} A_{1n}\\ A_{2n}  \end{array} \right)
\end{equation}
are proportional. This in turn means that we can transform them
all to zero, except for say
\begin{equation}
\left ( \begin{array}{c} A_{13}\\ A_{23} \end{array} \right),
\end{equation}
by going into a new basis for the fields
$\phi_k(\vec{p})$. So, finally, we have
transformed the formulation of the fields in such a way that only
the upper left three by three corner of the $A$ matrix is
non-zero.

This, in turn, means
that we can treat the bosonic model in the region of interest, by
studying the spectrum of a (generic) antisymmetric $3\times 3$
matrix with real elements, or rather such a matrix multiplied by
$i$. Let us immediately notice that such a matrix is parameterised
by \underline{three} parameters. The matrix and thus the spectrum,
to the accuracy we are after, can only depend on three of the
momentum parameters. In other words the dispersion relation will
depend trivially on all but 3 parameters in the linear
approximation. By this linear approximation, we here mean the
approximation in which the ``poor physicist'' can only work with a
small region in momentum parameter space also---not only in
energy. In this region we can trust the lowest order Taylor
expansion in the differences of the momentum parameters from their
starting values (where the nearest levels merge). Then the
$\omega$-eigenvalues---i.e.~the dispersion relation---will not
vary in the direction of a certain subspace of co-dimension three.
Corresponding to these directions the velocity components of the
described boson particle will therefore be zero! The boson, as
seen by the ``poor physicist'', can only move inside a three
dimensional space; in other directions its velocity must remain
zero. It is in this sense we say that the three-dimensionality of
space is explained!

The form of the equations of motion for the bosonic fields, in
this low excitation regime where one can use the lowest order
Taylor expansion in the momentum parameters, is also quite
remarkable: after a linear transformation in the space of
``momentum parameters'', they can be transformed into the {\em
Maxwell equations} with the fields being complex (linear)
combinations of the magnetic and electric fields. Let us now
derive this result.

The lowest order Taylor expansion of the antisymmetric matrix
$A_{ij}(\vec{p}, \vec{x})$ in $\vec{p}$, with $\vec{x}$-dependent
coefficients, takes the form:
\begin{equation}
\label{eq14c} A = \left ( \begin{array}{c|c|c}0 &
c^{3i}(\vec{x})p_i &
-c^{2i}(\vec{x})p_i\\
-c^{3i}(\vec{x})p_i & 0 & c^{1i}(\vec{x})p_i\\
c^{2i}(\vec{x})p_i & -c^{1i}(\vec{x})p_i & 0\\
\end{array} \right).
\end{equation}
Its eigenvalues are $-i\omega = 0,\pm
i\sqrt{g^{ij}(\vec{x})p_ip_j}$, where the real symmetric $3 \times
3$ spatial metric tensor is:
\begin{equation}
 g^{ij}_{Maxwell}(\vec{x})=
\sum_{t=1}^{3} c^{ti}(\vec{x})c^{tj}(\vec{x}).
\label{spatialmetric}
\end{equation}
Since, at the outset, our model is pre-Einsteinian in
the sense of having an absolute time, we can form a
$4 \times 4$ space-time metric $g^{\mu\nu}_{Maxwell}$,
by supplementing equation (\ref{spatialmetric}) with
the definitions:
\begin{equation}
g^{00}_{Maxwell} = 1 \qquad \hbox{and} \qquad
g^{0i}_{Maxwell} = g^{i0}_{Maxwell} = 0.
\label{temporalmetric}
\end{equation}

It is possible to construct the matrix square root
$\sqrt{g^{..}}^{ij}_{Maxwell}$ of the spatial metric tensor
(\ref{spatialmetric}) and,
by making an orthogonal transformation on the fields
$\phi_k(\vec{p})$, we can choose a basis in which $A_{ij}(\vec{p},
\vec{x})$ becomes:
\begin{equation}
\label{eq14g}
 A = \left (
\begin{array}{c|c|c}0 & \sqrt{g^{..}}^{3i}_{Maxwell}(\vec{x})p_i
& -\sqrt{g^{..}}^{2i}_{Maxwell}(\vec{x})p_i\\
-\sqrt{g^{..}}^{3i}_{Maxwell}(\vec{x})p_i & 0 &
\sqrt{g^{..}}^{1i}_{Maxwell}(\vec{x})p_i\\
\sqrt{g^{..}}^{2i}_{Maxwell}(\vec{x})p_i &
-\sqrt{g^{..}}^{1i}_{Maxwell}(\vec{x})p_i & 0\\
\end{array} \right).
\end{equation}
{\em A priori} the spatial metric is $\vec{x}$-dependent. However we
speculate that, if gravity can be incorporated into our theory,
the spatial metric becomes flat for some dynamical reason (cosmological
inflation?). Henceforth we assume that translational invariance
arises in this way and thus obtain:
\begin{equation}
\label{eq14}
A = \left ( \begin{array}{c|c|c}0 & p_3 & -p_2\\
-p_3 & 0 & p_1\\ p_2 & -p_1 & 0\\ \end{array} \right)
\end{equation}
in the chosen basis.

We can now make a Fourier transform of the three fields
$\phi_j(\vec{p})$ into the $\vec{x}$-representation. These new
position space fields $\phi_j(\vec{x})$ are no longer Hermitian.
However, it follows from the assumed Hermiticity of the
$\phi_j(\vec{p})$ that, in the $\vec{x}$-representation, the real
parts of the fields $\phi_j(\vec{x})$ are even, while the
imaginary parts are odd functions of $\vec{x}$:
\begin{equation}
\vec{\phi}(\vec{x}) = \vec{\phi}^{\dagger}(-\vec{x}).
\label{hermx}
\end{equation}
We now want to extract the real and imaginary parts and identify
them with the magnetic $\vec{B}$ and electric $\vec{E}$ fields.
This can be done in many ways, in as far as we could take
\begin{equation}
\label{ep} e^{i\delta}\vec{\phi}(\vec{x}) = \vec{B}(\vec{x}) + i
\vec{E}(\vec{x})
\end{equation}
with an arbitrary phase $\delta$. For definiteness we shall,
however, choose an especially nice identification with the
property that the field $\vec{\phi}(\vec{x})$ constructed
from $\vec{E}$ and $\vec{B}$ has simple transformation
properties under $P$, $T$ and $PT$:
\begin{equation}
P\vec{\phi}(\vec{x})P^{-1} = \vec{\phi}^{\dagger}(-\vec{x}),
\qquad
T\vec{\phi}(\vec{x})T^{-1} = \vec{\phi}(\vec{x})
\quad \hbox{and} \quad
PT\vec{\phi}(\vec{x})(PT)^{-1} = \vec{\phi}^{\dagger}(-\vec{x}).
\end{equation}
It then follows from (\ref{hermx}) that
\begin{equation}
PT\vec{\phi}(\vec{x})(PT)^{-1} = \vec{\phi}(\vec{x})
\quad \hbox{and} \quad
P\vec{\phi}(\vec{x})P^{-1} = \vec{\phi}(\vec{x}).
\label{PTe}
\end{equation}
These two expressions (\ref{PTe}) are, of course, equivalent
to each other, due to the relation $T\vec{\phi}(\vec{x})T^{-1}
=\vec{\phi}(\vec{x})$. Since under time reversal
\begin{equation}
\vec{E}(\vec{x}) \rightarrow -\vec{E}(\vec{x}),
\qquad \vec{B}(\vec{x}) \rightarrow \vec{B}(\vec{x}),
\qquad i \rightarrow -i,
\end{equation}
it is clear that to satisfy (\ref{PTe}), we should make the
identification
\begin{equation}
\vec{\phi}(\vec{x}) = \vec{B}(\vec{x}) +i\vec{E}(\vec{x})
\label{nice}
\end{equation}
as in~\cite{bled}.

Thus we identify the real and imaginary parts of
$\vec{\phi}(\vec{x})$ as magnetic and electric fields
$\vec{B}(\vec{x})$ and $\vec{E}(\vec{x})$ respectively:
$\phi_j(\vec{x}) = iE_j(\vec{x}) + B_j(\vec{x})$.
However the symmetry of these
Maxwell fields means that they must be in a configuration/state
which goes into itself under a parity reflection in the origin.
This is a somewhat strange feature which seems necessary for the
identification of our general fields with the Maxwell fields; we
discuss this feature further in section \ref{orbifold}.
For the moment let
us, however, see that we do indeed get the Maxwell equations in
the free approximation with the proposed identification.

By making the inverse Fourier transformation back to momentum
space, we obtain the following identification of the fields
$\phi_j(\vec{p})$ in our general quantum field theory with the
electric field $E_j(\vec{p})$ and magnetic field $B_j(\vec{p})$
Fourier transformed into momentum space:
\begin{equation}
\label{identification}
\left ( \begin{array}{c}\phi_1(\vec{p})\\
\phi_2(\vec{p})\\\phi_3(\vec{p})\\ \end{array} \right)=
\left ( \begin{array}{c}iE_1(\vec{p}) +  B_1(\vec{p})\\
iE_2(\vec{p}) + B_2(\vec{p})\\i E_3(\vec{p}) +B_3(\vec{p})\\
\end{array} \right).
\end{equation}
We note that the Fourier transformed electric field $E_j(\vec{p})$
in the above ansatz (\ref{identification}) has to be purely
imaginary, while the magnetic field $B_j(\vec{p})$ must be purely
real.

By using the above identifications, eqs.~(\ref{eq14}) and
(\ref{identification}), the equations of motion (\ref{eq8}) take
the following form
\begin{equation}
\left ( \begin{array}{c}i\dot{E}_1(\vec{p}) +  \dot{B}_1(\vec{p})\\
i\dot{E}_2(\vec{p}) + \dot{B}_2(\vec{p})\\i \dot{E}_3(\vec{p})
+ \dot{B}_3(\vec{p})\\
\end{array}
\right)  =\left ( \begin{array}{c|c|c}0 & p_3 & -p_2\\
-p_3 & 0 & p_1\\ p_2 & -p_1 & 0\\ \end{array} \right)
\left ( \begin{array}{c}iE_1(\vec{p}) +  B_1(\vec{p})\\
iE_2(\vec{p}) + B_2(\vec{p})\\i E_3(\vec{p}) +B_3(\vec{p})\\
\end{array} \right).
\end{equation}
We can now use the usual Fourier transformation identification in
quantum mechanics to transform these equations to the
$\vec{x}$-representation, simply from the definition of $\vec{x}$
as the Fourier transformed variable set associated with $\vec{p}$,
\begin{equation}
p_j  = i^{-1} \partial_j.
\end{equation}
Thus in $\vec{x}$-representation the equations of motion become
\begin{equation}
\left ( \begin{array}{c}i\dot{E}_1(\vec{x}) +  \dot{B}_1(\vec{x})\\
i\dot{E}_2(\vec{x}) + \dot{B}_2(\vec{x})\\i \dot{E}_3(\vec{x})
+\dot{B}_3(\vec{x})\\
\end{array}
\right) =\left ( \begin{array}{c|c|c}0 & -i\partial_3 & i\partial_2\\
i\partial_3 & 0 & -i\partial_1\\ -i\partial_2 & i\partial_1 & 0\\
\end{array}
\right)\left ( \begin{array}{c}iE_1(\vec{x}) +  B_1(\vec{x})\\
iE_2(\vec{x}) + B_2(\vec{x})\\i E_3(\vec{x}) +B_3(\vec{x})\\
\end{array} \right).
\end{equation}
The imaginary terms in the above equations give rise to the
equation
\begin{equation}
\label{iprop} \dot{\vec{E}}(\vec{x}) =
\mathrm{curl}\, \vec{B}(\vec{x}),
\end{equation}
while the real parts give the equation:
\begin{equation}
\label{realpart} \dot{\vec{B}}(\vec{x}) =
- \mathrm{curl}\, \vec{E}(\vec{x}).
\end{equation}
These two equations are just the Maxwell equations in the absence
of charges and currents, except that strictly speaking we miss two
of the Maxwell equations, namely
\begin{equation}
\label{missing} \mathrm{div}\, \vec{E}(\vec{x}) =0 \quad
\hbox{ {and} } \quad \mathrm{div}\, \vec{B}(\vec{x}) =0.
\end{equation}
However, these two missing equations are derivable from the other
Maxwell equations in time differentiated form. That is to say, by
using the result that the divergence of a curl is zero, one can
derive from the other equations that
\begin{equation}
\label{timederived}
\mathrm{div}\, \dot{\vec{E}}(\vec{x}) =0 \quad \hbox{
{and}} \quad \mathrm{div}\, \dot{\vec{B}}(\vec{x}) =0
\end{equation}
which is though not quite sufficient. Integration of the resulting
equations (\ref{timederived}) effectively replaces the $0$'s on
the right hand sides of equations (\ref{missing}) by terms
constant in time, which we might interpret as some constant
electric and magnetic charge distributions respectively. In our
free field theory approximation, we have potentially ignored such
terms. So we may claim that, in the approximation to which we have
worked so far, we have derived the Maxwell equations sufficiently
well. We shall consider interactions with the fermion field in
section \ref{Interaction}.

For later use, it is now convenient to introduce the usual
relativistic notation for the second rank
antisymmetric electromagnetic field $F^{\alpha\beta}$ tensor
\begin{equation}
F^{\alpha\beta} =
\begin{pmatrix}
0 & -E_1 & -E_2 & -E_3 \\
E_1 & 0 & -B_3 & B_2 \\
E_2 & B_3 & 0 & -B_1 \\
E_3 & -B_2 & B_1 & 0
\end{pmatrix}
=
\begin{pmatrix}
0 & -{\rm Im}\, \phi_1 & -{\rm Im}\, \phi_2 & -{\rm Im}\, \phi_3 \\
{\rm Im}\, \phi_1 & 0 & -{\rm Re}\, \phi_3 & {\rm Re}\, \phi_2 \\
{\rm Im}\, \phi_2 & {\rm Re}\, \phi_3 & 0 & -{\rm Re}\, \phi_1 \\
{\rm Im}\, \phi_3 & -{\rm Re}\, \phi_2 & {\rm Re}\, \phi_1 & 0
\end{pmatrix}
\end{equation}
and the corresponding dual tensor
\begin{equation}
{\cal{F}}^{\alpha\beta} = \frac{1}{2}
\epsilon^{\alpha\beta\gamma\delta} F_{\gamma\delta}
\qquad \mathrm{where} \qquad
F_{\gamma\delta} = g_{\alpha\gamma}F^{\alpha\beta} g_{\delta\beta}.
\end{equation}
Here $\epsilon^{\alpha\beta\gamma\delta}$ is the totally antisymmetric
symbol in four indices and $g_{\mu\nu} = g_{\mu\nu}^{Maxwell}$,
as defined in equations (\ref{spatialmetric}) and
(\ref{temporalmetric}) (still assuming flatness), is the usual metric
of special relativity. Thus we have:
\begin{equation}
{\cal{F}}^{\alpha\beta} =
\begin{pmatrix}
0 & -{\rm Re}\, \phi_1 & -{\rm Re}\, \phi_2 & -{\rm Re}\, \phi_3 \\
{\rm Re}\, \phi_1 & 0 & {\rm Im}\, \phi_3 & -{\rm Im}\, \phi_2 \\
{\rm Re}\, \phi_2 & -{\rm Im}\, \phi_3 & 0 & {\rm Im}\, \phi_1 \\
{\rm Re}\, \phi_3 & {\rm Im}\, \phi_2 & -{\rm Im}\, \phi_1 & 0
\end{pmatrix}
\end{equation}
By constructing the ``self-dual'' combination we get
\begin{equation}
F^{\alpha\beta} +i{\cal{F}}^{\alpha\beta} =
\begin{pmatrix}
0 & -i\phi_1^{\dagger} & -i\phi_2^{\dagger} & -i\phi_3^{\dagger} \\
i \phi_1^{\dagger} & 0 & -\phi_3^{\dagger} & \phi_2^{\dagger} \\
i \phi_2^{\dagger} & \phi_3^{\dagger} & 0 & -\phi_1^{\dagger} \\
i \phi_3^{\dagger} & -\phi_2^{\dagger} & \phi_1^{\dagger} & 0
\end{pmatrix}
\end{equation}
and clearly the conjugate ``anti self-dual'' combination is then
\begin{equation}
F^{\alpha\beta} -i{\cal{F}}^{\alpha\beta} =
\begin{pmatrix}
0 & i \phi_1 & i \phi_2 & i \phi_3 \\
-i\phi_1 & 0 & -\phi_3 & \phi_2 \\
-i\phi_2 & \phi_3 & 0 & -\phi_1 \\
-i\phi_3 & -\phi_2 & \phi_1 & 0
\end{pmatrix}.
\label{FmiFdual}
\end{equation}

\section{Derivation of the Weyl equation}
\label{Weylderivation}

Let us now turn to the application of the Jahn-Teller or
Homolumo-gap effect to a system of fermions in our general field
theory model. We shall assume that the Homolumo-gap effect turns
out to be strong enough to ensure that the Fermi-surface just gets
put to a place where the density of levels is very low. Actually
it is very realistic that a gap should develop in a field theory
with continuum variables $\vec{p}$ labelling the single particle
states. That is namely what one actually sees in an insulator;
there is an appreciable gap between the last filled {\em band} and
the first empty band. However, if the model were totally of this
insulating type, the poor physicist would not ``see'' anything,
because he is supposed to be unable to afford to raise a particle
from the filled band to the empty one. So he can only see
something if there are at least {\em some} fermion single particle
states with energy close to the Fermi-surface.

We shall now divide up our discussion of what happens near the
Fermi-surface according to the number of components of the fermion
field that are relevant in this neighbourhood.
Let us denote by $n$ the number
of fermion field components, which contribute significantly to
the eigenstates near the Fermi-surface in the small region
of momentum space we choose to consider.

We shall see that, instead of working with the antisymmetric matrix
with purely imaginary matrix elements $iA_{ij}$, near the
Fermi surface
we can work rather with asymmetric Hermitian
matrices. In fact we have already noted that the eigenvalues for
the matrix $iA_{ij}$ all the time occur together as pairs
$\pm \omega$. Thus, if at first we
want to consider matrices of the
antisymmetric form $iA_{ij}$ with levels near the Fermi surface,
we must also keep the corresponding ones
having eigenvalues nearly equal to minus the Fermi energy. So
if, after replacing the matrix by an effective one, we want
to work near the Fermi surface, we should keep
those eigenvalues in the matrix which are both near the
Fermi surface and near the opposite of the Fermi energy.
Thus, taking the number of relevant eigenvalues near the Fermi
surface to be $n$,
the number of rows and of columns to be kept in the antisymmetric
$iA_{ij}$ formalism should be $2n$. The matrix $iA_{ij}^{near
F.S.}$,  with which we then replace  $iA_{ij}$
near the Fermi surface, is a $2n$ by $2n$ matrix that is
very close to the matrix with the $n$-times degenerate  eigenvalue
 $\omega_{FS}$, i.e the Fermi energy, and with the other $n$
eigenvalues also collapsing to one $n$-times degenerate eigenvalue
$-\omega_{FS}$. Thus, to the crudest approximation, this matrix
$iA_{ij}^{near F.S.}$ is proportional to the matrix ${\bf J} $
consisting of four blocks, with the $n$ by $n$ unit matrix and
minus the $n$ by $n$ unit matrix in the off-diagonal blocks and
zero in the diagonal blocks. At a certain point $\vec{p}_0$ 
in momentum space
(or in phase space if we think in terms of phase space), we shall
in fact take the $n$ eigenvalues to be just the Fermi energy
$\omega_{FS}$ and the other $n$ to be $-\omega_{FS}$. When we move
a little away from this point, {\em a priori} terms linear in a
Taylor expansion around that point will appear all over the $2n
\times 2n$ matrix $i\left\{A_{ij}^{near F.S.}\right\} = i {\bf
A}^{near F.S.}$.
\begin{equation}
i {\bf A}^{near F.S.} \approx i\omega_{FS} {\bf J }
+ i\frac{\partial {\bf A}^{near F.S.}}{\partial p_i} (p_i -p_{0i})
= i\omega_{FS} \left(
\begin{array}{cc} \mathbf{0} & \mathbf{1} \\ \mathbf{-1} &
\mathbf{0} \end{array}\right) + i\frac{\partial {\bf A}^{near
F.S.}}{\partial p_i} (p_i - p_{0i})
\end{equation}
Parts of the linear term $ i\frac{\partial {\bf A}^{near
F.S.}}{\partial p_i} (p_i-p_{0i})$ which, as matrices, 
do not commute with
${\bf J}$ will only contribute at second order in perturbation
theory. So we shall ignore their effects and only include terms
contributing to $iA_{ij}^{near F.S.}$ {\em which have the property
of commuting with} {\bf J}. Such contributions are characterized
by having the two $n$ by $n$ diagonal blocks being identical and
the two off-diagonal blocks being equal except for an overall
sign. Since the matrix shall be of the same type as $iA_{ij}^{near
F.S.}$, i.e.~purely imaginary and antisymmetric, the $n$ by $n$
diagonal blocks are antisymmetric while the $n$ by $n$
off-diagonal blocks are symmetric. There are thus clearly $n^2$
free real variables. This is the same number as for an $n$ by $n$
Hermitian complex matrix. We can indeed easily make a
correspondence between the type of linear contributions to
$iA_{ij}^{near F.S.}$ which we have to consider and $n\times n$
Hermitian matrices. This linear contribution takes the form
\begin{equation}
 iA_{ij} = i\left(
\begin{array}{cc} A^{n\times n}_A & A^{n\times n}_S \\
-A^{n\times n}_S & A^{n\times n}_A
\end{array}\right) 
= A^{n\times n}_S \otimes \left(-\sigma^2\right)
+ A^{n\times n}_A \otimes i\sigma^0
\end{equation}
and the corresponding Hermitian matrix is given by
\begin{equation}
H= A^{n\times n}_S + i A^{n\times n}_A,
\label{Hmatrix}
\end{equation}
where $ A^{n\times n}_S$ and $A^{n\times n}_A$ denote symmetric
and antisymmetric matrices respectively. In this way we basically
interpret the matrix ${\bf J}$ as the imaginary element $i$. For
the purpose of calculating the positive eigenvalues $\omega$ near
$\omega_{FS}$, we can replace the purely imaginary antisymmetric
matrix $ i {\bf A}^{near F.S.}$ by the Hermitian matrix $H$ of
equation (\ref{Hmatrix}).

So now let us consider the
level-density near the Fermi-surface for $n$ complex fermion field
components.

\subsection{The case of $n=0$ relevant levels near the Fermi-surface}

The $n=0$ case must, of course, mean that there are no levels at
all near the Fermi-surface in the small momentum range considered.
This corresponds to the already mentioned insulator case. The
poor physicist sees nothing from such regions in momentum space
and he will not care for such regions at all. Nonetheless
this is the generic situation close to the Fermi surface and
will apply for most of the momentum space.

\subsection{The case of $n$ = 1 single relevant level
near the Fermi-surface}

In this case the generic situation will be that, as a certain
component of the momentum is varied, the level will vary
continuously in energy. This is the kind of behaviour observed
in a metal. So there will be a rather smooth density
of levels and such a situation is not favoured by the Homolumo gap
effect, if there is any way to avoid it.

\subsection{The case of $n$ = 2 relevant levels near the Fermi-surface}

In this situation a small but almost trivial calculation is
needed. We must estimate how a Hamiltonian,
described effectively as a 2 by 2 Hermitian matrix $H$ with
matrix elements depending on the momentum $\vec{p}$,
comes to look in the generic case---\i.e. when nothing is
fine-tuned---and especially how the level density behaves. That
is, however, quite easily seen, when one remembers that the three
Pauli matrices and the unit 2 by 2 matrix together form a basis
for the four dimensional space of two by two matrices. All
possible Hermitian 2 by 2 matrices can be expressed as linear
combinations of the three Pauli matrices $\sigma^j$ and the
unit 2 by 2 matrix $\sigma^0$ with real coefficients.
We now consider a linearized Taylor
expansion\footnote{A related discussion of the redefinition
of spinors has been given in the context of the low
energy limit of a Lorentz violating QED model \cite{Colladay}.}
of the momentum dependence of such matrices, by taking
the four coefficients to these four matrices to be arbitrary
linear functions of the momentum minus the ``starting momentum''
$\vec{p_0}$, where the two levels become degenerate with
energy $\omega(\vec{p_0})$.
That is to say we must take the Hermitian 2 by 2
matrix to be
\begin{equation}
H = \sigma^a V^i_a (p_i - p_{0i}) +
\sigma^0 \omega(\vec{p_0})
= \sigma^a V^i_a(\vec{x}) (p_i - A_i(\vec{x})) +
\sigma^0 A_0(\vec{x})
\label{Weyl}
\end{equation}
where we formally identify the $\vec{p}_0$ contributions
as a four-potental $A_a(\vec{x})=V^i_a(\vec{x})A_i(\vec{x})$.
This can actually be interpreted as the Hamiltonian for a
particle obeying the Weyl equation, by defining the ``covariant''
momenta
\begin{equation}
P_1 = V^i_1(\vec{x})(p_i - A_i(\vec{x})), \qquad
P_2 = V^i_2(\vec{x})(p_i - A_i(\vec{x})),
\qquad P_3 = V^i_3(\vec{x})(p_i - A_i(\vec{x})),
\end{equation}
and the renormalised Hamiltonian
\begin{equation}
H_{new} = H - \sigma^0 V^i_0(\vec{x})(p_i - A_i(\vec{x})) -
\sigma^0 A_0(\vec{x})
= \vec{\sigma}\cdot\vec{P},
\end{equation}
\begin{equation}
\omega_{new} = \omega - V^i_0(\vec{x})(p_i - A_i(\vec{x})) -
A_0(\vec{x}).
\label{Hnew}
\end{equation}
For simplicity, we shall here assume translational invariance and
take the $V_a^i(\vec{x})$, $A_i(\vec{x})$ and $A_0(\vec{x})$ to be
constants independent of position. The renormalisation of the
energy, eq.~(\ref{Hnew}), is then the result of transforming away
a constant velocity $V_0^i$ in $d$ dimensions carried by all the
fermions, using the change of co-ordinates $x^{\prime i} = x^i -
tV_0^i$, and measuring the energy relative to $\omega(\vec{p_0})$.
Note that the ``starting momentum" $\vec{p_0}$ will generically be
of the order of a fundamental (Planck scale) momentum, which
cannot be significantly changed by a ``poor physicist". So the
large momentum $\vec{p_0}$ effectively plays the role of a
conserved charge at low energy.

A trivial calculation
for the Weyl equation, $H_{new} \psi = \omega_{new} \psi$, leads
to a level density with a thin neck, behaving like
\begin{equation}
 \rho \propto \omega_{new}^2.
\end{equation}
According to our strong assumption about Homolumo-gap effects, we
should therefore imagine that the Fermi-surface in this case would
adjust itself to be near the $\omega_{new} =0$ level. Thereby there
would then be the fewest levels near the Fermi-surface.

\subsection{The cases $n\ge 3$}

For $n$ larger than $2$ one can easily find out that, in the
neighbourhood of a point where the $n$ by $n$ general Hamiltonian
matrix deviates by zero from the unit matrix, there are
generically branches of the dispersion relation for the particle
states that behave in the metallic way locally, as in the case
$n=1$. This means that the level density in such a neighborhood
has contributions like that in the $n=1$ case, varying rather
smoothly and flatly as a function of $\omega$. So these cases are
not so favourable from the Homolumo-gap point of view.

\subsection{Conclusion
 of the various $n$ cases for the fermion model}

The conclusion of the just given discussion of the various
$n$-cases is that, while of course the $n=0$ case is the ``best''
case from the point of view of the Jahn-Teller or Homolumo-gap
effect, it would not be noticed by the ``poor physicist'' and thus
would not be of any relevance for him. The next ``best'' from the
Homolumo-gap point of view is the case $n=2$ of just two complex
components (corresponding to 4 real components) being relevant
near the Fermi-surface. Then there is a neck in the distribution
of the levels, which is not present in the cases $n=1$ and $n>2$.

So the ``poor physicist'' should in practice observe the case
$n=2$, provided the Homolumo-gap effect is sufficiently
strong (a perhaps suspicious assumption).

Now, as we saw, this case of $n=2$ means that the fermion field
satisfies a Weyl equation, formally looking like the Weyl equation
in just 3+1 dimensions! It should however be noticed that there
are indeed more spatial dimensions, by assumption, in our model.
In these extra spatial dimensions, the fermions have the same
constant velocity which we were able to renormalise to zero, because
the Hamiltonian only depends on the three momentum
components $\vec{P}$ in the Taylor expandable region accessible to
the ``poor physicist''. The latter comes about because there are
only the three non-trivial Pauli matrices that make the single
particle energy vary in a linear way around the point of expansion.
In this sense the number of spatial dimensions comes out as equal to
the number of Pauli matrices.

\section{Charge conservation}
\label{chargeconservation}

We should like to remark here that the existence of a dispersion
relation for a type of particle
giving the energy as a function of momentum, which we now assume
to be conserved, can very easily lead to the effective
conservation of the particle number for the ``poor physicist''.
The point is simply the following:

The poor physicist who has only very little energy at his disposal
can normally only work with -- i.e. produce or find manageable
particles in -- a very tiny part of the momentum range. If he seeks
to go outside the very small range he will need bigger energy. By
manageable we here mean that he can accelerate the particles in
the momentum range in question, so that a filled Dirac sea is
useless unless he can afford to bring the particles out of it. So
only the part of the Dirac sea very close in energy to the
Fermi surface is considered manageable. But now, generically, the
tiny range of momentum with which he can work would, in a
non-rotational invariant theory of the type we consider here, almost
certainly not just be for zero momentum. But we would rather
expect some
momentum range which is small in extension, compared to the
fundamental scale of momentum, but surrounding a momentum value
typically of order unity in the supposed very large fundamental
scale unit.

This really means that all the particles, with which the poor
physicist can work, have one huge momentum of order
the fundamental scale plus a tiny little bit identified as the
momentum he talks about in the laboratory. But this situation now
has the consequence that in practice the particle number cannot
change! Such a change would require that the big momentum of
fundamental scale should be dispensed with in order to, say,
destroy the particle. This will normally be very difficult, even
with some violation of momentum conservation, since it would have
to be a very special amount of momentum that should be dispensed
with and there would be very little phase space for that to happen.
With momentum
conservation assumed, it becomes of course impossible unless the
poor physicist can afford to build up to the huge momentum;
he cannot do so in practice if, say, the huge momentum is of order
the Planck momentum and he needs to transfer it onto a
single or very few particles.

We can therefore say that our type of model gives a natural
mechanism for generating a conserved particle number charge.

\subsection*{Avoiding particle anti-particle level repulsion}
\label{cpt}

That there are such charge conservation mechanisms for the
particles, which finally become relativistic particles, is
actually quite needed for the following reason having to do with
the CPT-theorem:

Both for the Weyl fermion and for the photon we have, due to CPT
symmetry in a final relativistic theory,
an exact degeneracy between a particle and
an anti-particle having the same momentum. Suppose now
the theory were quite generic and everything not forbidden
should occur as terms in the Hamiltonian or commutators,
and we had no selection rules like those arising from
rotation symmetry at first. Then the particle and the
anti-particle could mix and there would generically be a level
repulsion, driving the energies of the
two states with the same momentum away from
each other! But that would be catastrophic for reproducing a
theory satisfying the CPT-theorem.

However, as explained above, we indeed do have a charge
conservation in our model for
both the complex $\vec{\phi}(\vec{x})$ field and the Weyl field.
In the boson case, it turns out to be the duality charge
corresponding to the generator of duality transformations
of the electromagnetic fields:
\begin{equation}
\vec{B} \longrightarrow \vec{B} \cos\theta + \vec{E} \sin\theta,
\qquad
\vec{E} \longrightarrow \vec{E} \cos\theta - \vec{B} \sin\theta.
\end{equation}
So we have a chance, at least, of
obtaining totally degenerate CPT-connected states and thus a
chance to get the final theory of relativity out.
According to the potential problems associated with level repulsion,
we should not have been able to obtain Majorana particles or bosons
not having a charge or particle number.
Otherwise we would have expected to generically obtain
a CPT non-conserving theory.

So we can say that if one should find Majorana particles, except if
they are due to some tiny correction starting from particles
with a conserved
particle number, it would be in contradiction with our present
random dynamics model. Especially supersymmetry models,
with lots of Majorana fermions, look a bit dangerous from the
point of view of our random dynamics model.

\section{Interaction of Weyl and photon fields}
\label{Interaction}

We shall now argue that it is indeed quite natural to obtain
the well-known quantum electrodynamics, when we allow for
interactions. It
is a well-known problem for electrodynamics that there is
essentially no nice way to write down the interaction Lagrangian
term -- usually taken to be of the form $
A_a\psi^{\dagger}\sigma^a\psi$ -- without
the use of the electromagnetic {\em four potential} $A_a$
or $A_{\mu}$. One can in principle, in some gauge, reconstruct the
$A_{\mu}$-field from an integral over the $\vec{E}$ and $\vec{B}$
fields. Such constructions will turn out to be non-local (in the
sense that you integrate over different space-time regions)
and rather inelegant. However above, we have
managed to obtain the free Maxwell equations without
using the $A_{\mu}$ field at any stage.

Let us recall that we introduced a field $A_a$ in equation
(\ref{Weyl}). That there should be such a field $A_a$ is
easily seen, by considering the most general linear equation
of motion for a fermion with two components. That is to say we
shall look at a Weyl-like equation or
Schr\"{o}dinger-like equation with \underline{two} components:
\begin{equation}
i\frac{d}{dt}{\psi_1 \choose \psi_2}
= -\sum_a \sigma^a i D_a {\psi_1 \choose \psi_2}
\label{motion}
\end{equation}
This form of the equation will be the most general Hermitian
Hamiltonian Schr\"{o}dinger equation, if $iD_a$ are four general
Hermitian operators in the geometrical, i.e.~$\vec{x}-\vec{p}$,
degrees of freedom. As above, we shall take the philosophy that,
for the poor physicist, the momentum range can only be small
compared to the fundamental scale, while $\vec{x}$ may be large.
Thus we Taylor expand $iD_a$ in $\vec{p}$, while each term has an
arbitrary $\vec{x}$ dependence:
\begin{equation}
-iD_a = A_a(\vec{x}) + \vec{p}\cdot \vec{V}_a(\vec{x}).
\label{Da}
\end{equation}
The notation here is suggestive of the physical interpretation we
are suggesting. The dominant term in the Taylor expansion
$A_a(\vec{x})$ is to be interpreted as the four-potential.
However the index being $a$ suggests that it is a so-called
flat index, which means that it is associated with a vierbein,
$V^{\mu}_a$, formulation of general relativity, rather than
the coordinate or Einstein index $\mu$ on $A_{\mu}$, i.e.~really
$A_a=V_a^{\mu}A_{\mu}$. The next term
$\vec{p}\cdot \vec{V}_a(\vec{x})$ is essentially just the
vierbein we have to accept as coming out of the ``random"
momentum dependence of the fermion Hamiltonian around the two
state degenerate point as we considered it above.
Actually the Weyl field related vierbein $V^{\mu}_a$ is the
extension of $V^i_a$ in equation (\ref{Da}), by the further
components $V^0_a = \delta^0_a$ representing the left hand side
of equation (\ref{motion}). This is discussed further below.
Note that {\em a priori} the metric
associated with the Maxwell field is different to the Weyl
metric and, thus, this vierbein $V^{\mu}_a$ will not match with the
Maxwell metric.

By second quantisation of the above single fermion Hamiltonian,
we obtain the following second quantised Hamiltonian
\begin{equation}
\int \psi^{\dagger}(H_F + H_{INT})\psi \ d^d\vec{x}   =
\int \psi^{\dagger}(V_a^{\mu}p_{\mu} + A_a)\sigma^a\psi
\ d^d\vec{x}.
\end{equation}
So we can say that we have argued for an interaction term
\begin{equation}
H_{INT} = \int A_a\psi^{\dagger}\sigma^a\psi \ d^d\vec{x}
\label{interaction}
\end{equation}
as natural in $\vec{x}$-representation.

\subsection{The interacting model}

The philosophy is that we simply add this interaction term to the
free particle approximation boson and fermion Hamiltonians. But
now we remember that, for the bosons, we have relegated the
complications into the commutators. It is therefore the
commutators of the $\vec{\phi}$-fields and the $A_a$-fields which
may contain such possible complicated dynamics. It is also only
this commutator which can identify or create a ``same position''
relation between the space in which the Weyl fermion exists and
the one in which the Maxwell equation photon exists. Indeed it is
good to have in mind that, in the free approximation used up to
now, there is absolutely no sense in saying that a boson and a
Weyl fermion are at the same place. They live so to speak in
completely different geometrical spaces. Even when we developed
the interaction term (\ref{interaction}), the position variable in
the field $A_a(\vec{x})$ (later to be the electromagnetic
4-potential) is a position in the Weyl fermion geometrical space
and we cannot yet identify this position variable with that for
the Maxwell photon. Thus the identification of these two totally
different geometrical spaces comes about only via the {\em
commutators} of the $A_a$ and the $\vec{\phi}$ fields. In our
``perturbative" approach of considering the dominant terms in the
sense of having fewest field operator factors, we approximate the
commutator of $\vec{\phi}$ and $A_a$ as a c-number times a
$\delta$-function
\begin{equation}
[\phi_i(\vec{x}),A_a(\vec{x^{\prime}})] = F_{ia}(\vec{x})
\delta(\vec{x} - \vec{x^{\prime}}).
\end{equation}

Regarding the background fields $F_{ia}(\vec{x})$ and
$\vec{V}_a(\vec{x})$, we may take either of the two
following attitudes
\begin{enumerate}
\item
They are gravitational fields and depend on the dynamical
``rest'' variables.
\item
They are background fields which act like coupling constants.
\end{enumerate}
In the first case, we might hope one day to incorporate gravity
into our model and we could then treat them as constant on the
grounds of the gravitational field being weak and, hence,
approximately taking the trivial form $g^{\mu\nu}= \eta^{\mu\nu}$
and $V^{\mu}_a = \delta^{\mu}_a$. In the second case,
we could assume them to be constant on the grounds of simply
putting translational invariance into our model {\it ab initio}.

We shall see now that, basically from the consistency of the
equations of motion, we can find the form of the constant $F_{ia}$
in this commutator up to a constant factor $e$ being the electric
charge of the fermion. In fact the equation of motion for
the  Maxwell fields reads
\begin{equation}
\dot{\phi_i} = i[H,\phi_i] = i[H_B+H_{INT},\phi_i] =
i\epsilon_{ikl}\frac{\partial}{\partial x_k}\phi_l(\vec{x}) -
iF_{ia} \psi^{\dagger}(\vec{x}) \sigma^a \psi(\vec{x}) .
\label{eqmint}
\end{equation}
Here we used the notation of the coordinates associated
with the metric for the Maxwell fields, so that the
free Maxwell equations are without explicit vierbeins
or metrics. In this formulation, however, the Weyl
particle then needs a vierbein and the expression for
the conservation of the Weyl fermion number becomes
\begin{equation}
\partial_{\mu}j^{\mu} = \partial_{\mu} \left(
V^{\mu}_a j^a\right)
\end{equation}
where
\begin{equation}
j^a = \psi^{\dagger} \sigma^a \psi
\end{equation}
and $V^{\mu}_a$ is the constant vierbein.

\subsection{Constraining the commutator coefficients
$F_{ia}$ by consistency of equations of motion}

By taking the divergence of equation (\ref{eqmint}), it follows
that
\begin{equation}
\frac{d}{dt} \mathrm{div}\, \vec{\phi}(\vec{x}) = -i\partial_i
\left(F_{ia} \psi^{\dagger}(\vec{x}) \sigma^a
\psi(\vec{x})\right), \label{divphidot}
\end{equation}
showing that the time derivative of the divergence of
$\mathrm{div}\, \vec{\phi}$ only depends on the matter or fermion
fields. If we imagine that this $\mathrm{div}\, \vec{\phi}$ can be
expressed as a function of all the fields of the theory, we may
expand the time derivative
\begin{equation}
\mathrm{div}\,\dot{\vec{\phi}} =
\frac{\partial \mathrm{div}\, \vec{\phi}}
{\partial \vec{\phi}} \dot{\vec{\phi}}
+ \frac{\partial \mathrm{div}\, \vec{\phi}}
{\partial \psi_i} \dot{\psi_i}
\end{equation}
by differentiating with respect to the various fields.
Since, according to the equation of motion for $\psi$, even
$\dot{\psi}$ at first only contains $A_a$ and $\psi$ but no
$\vec{\phi}$ field, it is indicated that
$\mathrm{div}\, \vec{\phi}$ should not depend on the
$\phi_i$'s but rather on $\psi$ and perhaps
on $A_a$. Ignoring possible dependence on $A_a$,
we want the
time derivative of $\mathrm{div}\, \vec{\phi}$ to be
a quadratic expression in $\psi$. Then, since the equation of motion
leaves $\dot{\psi}$ linear in $\psi$, we really need
$\mathrm{div}\, \phi$ to be
a quadratic expression in $\psi$.

We are therefore driven to make an ansatz of the form
\begin{equation}
\mathrm{div}\, \vec{\phi} = \psi^{\dagger}\cal{O} \psi
\label{ansatz}
\end{equation}
where to keep fermion number conservation, we should have
the form with one $\psi^{\dagger}$ and one $\psi$.
We expand the operator $\cal{O}$, which is constant (in space), as
\begin{equation}
{\cal{O}} = W_a \sigma^a.
\label{Oform}
\end{equation}
Note that, since $\mathrm{div}\,\phi(\vec{x})$ is not Hermitian in
$\vec{x}$-representation, it is possible that the four numbers
$W_a$ are complex.

Now let us use the equation of motion for the fermion fields
(\ref{motion}), so as to compute the time derivative of our ansatz
(\ref{ansatz})
\begin{equation}
\frac{d}{dt} \mathrm{div}\, \vec{\phi}= -\frac{\partial
\psi^{\dagger}} {\partial \vec{x}} \vec{V}_a \sigma^a
W_b\sigma^b\psi -\psi^{\dagger}W_b\sigma^b \vec{V}_a \sigma^a
\frac{\partial \psi}{\partial \vec{x}}, \label{eq105}
\end{equation}
where we used the Hermitian conjugate of equation (\ref{motion})
without the $A_a$ term
\begin{equation}
i\frac{d}{dt} \psi^{\dagger} = -i \frac{\partial \psi^{\dagger}}
{\partial \vec{x}} V_a \sigma^a.
\end{equation}
By rearrangement of the terms in equation (\ref{eq105}), we
further obtain
\begin{equation}
\frac{d}{dt} \mathrm{div}\, \vec{\phi} =
-\frac{\partial}{\partial \vec{x}} \left(\psi^{\dagger}
\frac{1}{2}[\vec{V_a}\sigma^a, W_b\sigma^b]_+\psi\right)
- \psi^{\dagger}\frac{1}{2}[W_b\sigma^b,
\vec{V}_a\sigma^a]
\frac{{\stackrel{\leftrightarrow}{\partial}}}
{\partial \vec{x}}\psi
\label{divphidot2}
\end{equation}
According to equation (\ref{divphidot}) we do not want a term of
the second type with the derivative
$\frac{{\stackrel{\leftrightarrow} {\partial}}}{\partial \vec{x}}$
acting both ways, but only a term of the first type, where the
$\frac{\partial}{\partial \vec{x}}$ differentiation acts outside
the $\psi^{\dagger}.... \psi$ product. The requirement that the
second term disappears is that for all $i=1,2,3$ we have
\begin{equation}
W_bV^i_a\epsilon^{bac} = 0
\end{equation}
So in the case  -- normally true -- that the $3 \times 3$ matrix
$(V^i_a)_{a=1,2,3}$ is non-degenerate, we conclude
\begin{equation}
W_b = 0 \qquad \hbox{for} \ b=1,2,3.
\label{W123E0}
\end{equation}

In order that the first term in equation (\ref{divphidot2})
should agree with equation (\ref{divphidot}) we need
\begin{equation}
\sigma^aF_{ia}=\frac{1}{2i}[V^i_a\sigma^a,W_b\sigma^b]_+
\end{equation}
which in turn gives
\begin{equation}
iF_{i0} = \sum_{b=0}^3V^i_bW_b = W_0V^i_0
\label{FVtime}
\end{equation}
and
\begin{equation}
iF_{ia} = W_0V^i_a \qquad \hbox{for} \ a\neq 0.
\label{FVspace}
\end{equation}
But this means that we only get consistent equations when $F_{ia}$
takes this special form. Notice that, in our Hamiltonian (or
absolute time) presentation of the fermion theory, we did not use
a full 16 component vierbein, but only the 12 components with the
Einstein index being $i=1,2,3$. We did not need the components
$V^0_a$. Although we already introduced the further components in
the first part of this section, we shall formally re-introduce
them here. We do that by re-writing the expression (\ref{ansatz})
as
\begin{equation}
\mathrm{div}\, \vec{\phi} = W_0\psi^{\dagger}V^0_a\sigma^a\psi
\label{intrV0}
\end{equation}
where $W_0$ can be considered a complex charge, as we also do for
the following expression:
\begin{equation}
iF_{ia}\psi^{\dagger}\sigma^a\psi=
W_0\psi^{\dagger}V^i_a\sigma^a\psi.
\end{equation}
Here we used equations (\ref{FVtime}) and (\ref{FVspace}).
Consistency of equation (\ref{intrV0}) with equations
(\ref{ansatz}) and (\ref{Oform}), requires
\begin{equation}
W_a=W_0V^0_a.
\label{Wa}
\end{equation}
Combined with equation (\ref{W123E0}) this of course gives
\begin{equation}
V^0_0 =1, \qquad V^0_b =0, \quad \hbox{for}\ b\neq 0.
\label{4bein}
\end{equation}

Since our Weyl equation of motion was simply (\ref{motion}), we
see that it can indeed be precisely expressed by means of a four
dimensional vierbein of just the form to which we arrived, namely
(\ref{4bein}) and the already needed 12 components $\vec{V}_a$ in,
say, equation (\ref{Da}). In fact the equation of motion
(\ref{motion}) takes the following four dimensional form
\begin{equation}
V^{\mu}_a \sigma^a i D_{\mu} \psi = 0
\end{equation}
where now
\begin{equation}
D_{\mu}=\left(V^{-1}\right)_{\mu}^a
\left(D_a + \delta^0_a\frac{d}
{dt}\right)
=\left(V^{-1}\right)_{\mu}^a
\left(iA_a+ip_iV^i_a+\delta^0_a
\frac{d}{dt}\right) =
iA_{\mu} + \frac{\partial}{\partial x^{\mu}}
\end{equation}
and $A_{\mu}$ is defined so that
\begin{equation}
A_a=V^{\mu}_a A_{\mu}.
\end{equation}
From such a vierbein form of the Weyl equation, one immediately
finds that the conserved 4-current is
\begin{equation}
j^{\mu}=V^{\mu}_a\psi^{\dagger}\sigma^a\psi
\label{jmu}
\end{equation}
obeying
\begin{equation}
\partial_{\mu}j^{\mu}=0.
\end{equation}

From equations (\ref{FVtime}) and (\ref{FVspace}), we have
\begin{equation}
iF_{ia}=W_0V^i_a.
\end{equation}
So the equation of motion (\ref{eqmint}) becomes
\begin{equation}
\dot{\phi}_i(\vec{x}) = i\epsilon_{ikl} \frac{\partial}{\partial
x_k}\phi_l(\vec{x}) - W_0j^i(\vec{x})
\end{equation}
where $j^i$ is one of the spatial components of (\ref{jmu}).
We must admit that it looks strange here to have put an upper index
on $j^i$; it is a consequence of our notation for the vierbein
$V^{\mu}_a$ with an upper Einstein index $\mu$. Using
the space-time formulation (\ref{FmiFdual}), this equation can be
written as the spatial components for the covariant-looking
equation
\begin{equation}
\partial_{\alpha} (F^{\alpha\beta} -
i{\cal{F}}^{\alpha\beta})= -iW_0j^{\beta},
\label{em51}
\end{equation}
i.e. we got it for $\beta=i, \ i=1,2,3$. For $\beta = 0$
equation (\ref{em51}) corresponds to equation (\ref{ansatz}).
Taking the Hermitian conjugate of this equation gives us
\begin{equation}
\partial_{\alpha} (F^{\alpha\beta} +
i{\cal{F}}^{\alpha\beta})= iW_0^*j^{\beta},
\label{em52}
\end{equation}
since $j^{\beta}$, $F^{\alpha\beta}$ and $\cal{F^{\alpha\beta}}$
are all Hermitian operators.

Adding and subtracting,
we obtain
\begin{equation}
\partial_{\alpha}F^{\alpha\beta} = \frac{1}{2}
\left(iW_0^*-iW_0\right)j^{\beta}
\label{M1}
\end{equation}
and
\begin{equation}
\partial_{\alpha} {\cal{F}}^{\alpha\beta}  =
\frac{1}{2}
\left(W_0+W_0^*\right)j^{\beta}
\label{M2}
\end{equation}
which shows that the Weyl fermion has come to function as
a dyon with electric $e$ and magnetic $g$ charges given by
\begin{equation}
e= {\rm Im}\, W_0, \qquad g= {\rm Re}\, W_0.
\end{equation}
Although these equations look completely Lorentz covariant, we
still actually have a Lorentz non-invariant theory, because the
current $j^{\beta}(\vec{x})$ is given by (\ref{jmu}) and contains
the vierbein $V^{\beta}_a$ which is constant in space and time. We
can make this very important flaw manifest, by writing equations
(\ref{M1}) and (\ref{M2}) with the vierbein explicitly exposed:
\begin{equation}
\partial_{\alpha}F^{\alpha\beta}(\vec{x})=
eV^{\beta}_a\psi^{\dagger}(\vec{x})\sigma^a\psi(\vec{x}),
\qquad \partial_{\alpha}{\cal{F}}^{\alpha\beta} =
gV^{\beta}_a\psi^{\dagger}(\vec{x})\sigma^a\psi(\vec{x}).
\label{5354}
\end{equation}
Thus, even with our optimistic speculations, we did not obtain
full Lorentz invariance at first.

\section{Transforming away a dyon}

As already mentioned in section \ref{Maxwelleq},
the extraction of $\vec{E}$ and $\vec{B}$ from
$\vec{\phi}$ can be made in different conventions, provided
one does not require that the condition
$\vec{\phi}(\vec{x}) = \vec{\phi}^{\dagger}(-\vec{x})$
from the Hermiticity of the momentum representation field
be just represented as invariance of the field
{\em configuration} under P and PT.
Liberated from such requirements we may use equation (\ref{ep}),
which has an extra constant phase factor
$e^{i\delta}$ put on $\vec{\phi}(\vec{x})$, in order to
extract $\vec{E}$ and $\vec{B}$.

With this extraction of $\vec{B} + i\vec{E}$ we get
\begin{equation}
\partial_{\alpha}\left(e^{-i\delta}\left(F^{\alpha\beta}
-i{\cal{F}}^{\alpha\beta}\right)\right) =
-iW_0V^{\beta}_aj^a
\end{equation}
and
\begin{equation}
\partial_{\alpha}\left(e^{i\delta}\left(F^{\alpha\beta}
+i{\cal{F}}^{\alpha\beta}\right)\right) =
iW_0^*V^{\beta}_aj^a
\end{equation}
instead of equations (\ref{em51}) and (\ref{em52}). These revised
equations are of the usual form (\ref{5354}) but now with
\begin{equation}
e= {\rm Im}\left(W_0e^{i\delta}\right), \qquad
g= {\rm  Re}\left(W_0e^{i\delta}\right).
\end{equation}
With this phase $\delta$ available as a choice of interpretation,
we can arrange {\em for a single fermion field} that, for example,
the magnetic charge $g$ is zero. In this way we can achieve
conventional quantum electrodynamics {\em except that we still
have the non-trivial vierbein}. This means that the ``electron''
has its own metric tensor, which is different from that of the
Maxwell fields.

\section{The orbifold nature of the Universe: an extra prediction}
\label{orbifold}

In \cite{bled}, where we assumed translational invariance at the
outset, the space on which the Maxwell fields were defined was
indeed an orbifold. The orbifold identification in this case
corresponds to identifying the positions at $\vec{x}$ and at
$-\vec{x}$. This simple orbifold identification corresponds to the
reflection symmetry relations in phase space (orbifolding of phase
space) discussed in section \ref{twopointkernels} of this paper.
However, we have replaced the Weyl fields by effectively complex
fields and we do not have a corresponding symmetric state (of the
fermion fields). Therefore we effectively get the fermion fields
{\em doubled}, when we live on the orbifold. Suppose the Weyl
fields, which are genuinely near $\vec{x}$, are left-handed. Then
it is not difficult to see that the fields (also left-handed of
course) near -$\vec{x}$ will appear as right-handed to the
physicist at $\vec{x}$. Thus the effect of this reflection
symmetry or orbifold structure is really that the Weyl fermions
get doubled to become Dirac fermions. This poses a
phenomenological problem for our model, in as far as weak
interaction charges are not symmetric under left going to right.
So such a vectorlike symmetry prediction from the orbifold feature
will prevent an extension of our model to include the weak
interactions, unless we can somehow provide a crutch to cure this
problem.

With this reflection identification $\vec{\phi}(\vec{x})
=\vec{\phi}^{\dagger}(-\vec{x})$ we do not have locality, as was
tacitly assumed in the derivation of the Maxwell equations
(\ref{em51}) and (\ref{em52}), and thus these equations are
strictly speaking wrong. Rather we must consider together the
contributions to the Hamiltonian around a point $\vec{x}$ and the
reflected point -$\vec{x}$. Effectively we may ``bring'' the
interaction of the Maxwell field with the Weyl fermion field at
-$\vec{x}$ over to the point $\vec{x}$, by means of a reflection
symmetry operation P or PT. Really it is easy to see that we can
describe the effect of the interaction of the $\vec{\phi}(-x)$
field with the {\em a priori} left Weyl field -- in the region
around -$\vec{x}$ -- as an interaction of the field near
$\vec{x}$, i.e.~$\vec{\phi}(\vec{x})$, with a right-handed partner
field, giving simply an extra term on the right-hand side of the
Maxwell equations.

The degrees of freedom of the fermion near $-\vec{x}$ can, in
practice, be considered quite independently of those near
$\vec{x}$ and thus are best described by completely separate
fermionic fields, marked say by an index $R$ to remind us that
they now represent a right-handed Weyl particle. The ones at
$\vec{x}$ could then be marked analogously by an index $L$. The
vierbein $V^{\mu}_{Ra}$ for the $\psi_R$ degrees of freedom are
obtained by the parity operation on the original vierbein for the
{\em a priori} left Weyl field $V^{\mu}_a$. That is to say
\begin{equation}
V_{Ra}^{\mu} = - (-1)^{\delta^{\mu}_0}V^{\mu}_a,
\qquad V_{La}^{\mu} = V_a^{\mu}.
\end{equation}

The full Maxwell equations for the doubled theory
become
\begin{equation}
\partial_{\alpha}F^{\alpha\beta}=
eV_{La}^{\beta}\psi^{\dagger}_L\sigma^a\psi_L
+eV_{Ra}^{\beta}\psi^{\dagger}_R\sigma^a\psi_R
\end{equation}
and \begin{equation}
\partial_{\alpha}{\cal{F}}^{\alpha\beta} =
gV_{La}^{\beta}\psi^{\dagger}_L\sigma^a\psi_L
+gV_{Ra}^{\beta}\psi^{\dagger}_R\sigma^a\psi_R.
\end{equation}
This means that the Weyl fermion components R and L {\em a priori}
each have their own vierbein and thereby also their own metric
tensor. However, when we remember that due to the absolute time we
have $V_i^0=0$ from equation (\ref{4bein}), we see that with
\begin{equation}
g^{\mu\nu}_R = \eta^{ab}V_{Ra}^{\mu}V_{Rb}^{\nu},
\qquad g_L^{\mu\nu} = \eta^{ab}V_{La}^{\mu}V_{Lb}^{\nu},
\end{equation}
we have in fact
\begin{equation}
g_L^{\mu\nu} =g_R^{\mu\nu},
\end{equation}
because the sign from the parity reflection is squared in the
metrics. Thus it actually turns out that the whole Dirac fermion
-- here massless -- behaves in a Lorentz invariant way by itself.
However the  electromagnetic fields have a different metric.

\section{Conclusion and outlook}
\label{conclusion}

We started from a very general --- essentially random --- quantum
field theory, with the second quantized field operators only
defined as functions of general coordinates $ \vec{x}_{pre}$. The
interpretation of these general coordinates, in terms of physical
position and momentum, was then made in a way depending on the
Hamiltonian and commutator functions. This required a drastic
transformation of the single particle degrees of freedom, in order
to reach the identification with known physics -- meaning quantum
electrodynamics. The procedure was based on Taylor expanding in
the low energy and corresponding deviations in what we interpreted
as momentum, so that it only works in the low energy limit
relevant for the so-called ``poor physicist''. We obtained the
$3+1$ dimensional Maxwell equations and Weyl equations in the free
approximation, as already done in an earlier article \cite{bled}
under the assumption of translational invariance.

Furthermore we generalised the model to include a general
interaction between the fermions and the Maxwell fields. We also
made an attempt to even derive translational invariance similarly.
The latter attempt was less successful, although there is some
hope to succeed by including gravity.

\subsection{The great speculation that gravity may solve our remaining
problems!}

We did not really succeed fully in deriving
quantum electrodynamics from a
completely random or complicated theory in the low energy limit.
But we actually came so far that there is a hope of reaching this
goal, if somehow the gravity fields of general relativity emerge
from our approach (or are simply added by assumption) and the
background fields $\sqrt{g^{..}}^{ij}_{Maxwell}$ and $V_a^{\mu}$
are identified with the gravitational fields.

By speculatively incorporating gravity into the model in this way,
we may be able to resolve some of the major problems still
remaining in our model:

a) The problem of there still {\em not being translational
invariance} would be solved by the background fields depending on
the position, $\sqrt{g^{..}}^{ij}_{Maxwell}$ and $V_a^{\mu}$, now
being interpreted as gravitational fields. Then the lack of
translational invariance would just be interpreted as the effect
of gravitational fields being present, which are not translational
invariant. Thus the question of why we have translational
invariance would be brought back to the question of why the
gravitational fields have arranged themselves in such a way that
the space-time is so flat as to have translational invariance
approximately.

b) The rudiments of the {\em extra dimensions} must not disturb
our results, in order that our model be truly viable. The three
spatial dimensions and the one time dimension are singled out by
the velocity of the particles being of order unity in these
directions, while the velocity in the extra directions relative to
the 3 spatial dimensions is zero. If we were optimistic then we
could assume that the gravity theory would, in an analogous way,
single out three dimensions as the only ones in which rapid
particle motion could take place. This would mean that the
gravitational Lagrangian would lack the gradient squared terms
corresponding to the gradients in the {\em extra} directions. In
order to see what this could mean, we may think of the analogous
thing happening for an ordinary gauge field:

In \cite{layer} we studied a lattice gauge theory in a phase in
which there was, crudely speaking, confinement in some directions
-- dimensions -- while having an ordinary electrodynamics-like
phase in say four dimensions. This was achieved by having a strong
lattice gauge coupling  -- meaning a small coefficient on the
lattice action plaquette terms -- in the extra dimensions, while
having a weaker one in the normal dimensions. In this phase a
gauge field charge could not propagate from one ``layer'' to the
next, meaning it could not propagate in the dimensions with strong
coupling behaviour but would be confined to its layer. In the
suggested analogy, the analogue of the charge would be the energy
and momentum in gravity theory. Then having zero velocity particle
motions in the more than $3+1$ dimensions, which means zero
gradient term coefficient in these directions, is in turn
analogous to the strong coupling plaquette, i.e.~the plaquette
with a small coefficient on it. So we would expect a strong
coupling behaviour with momentum and energy confined in the more
than $3+1$ dimensions. This would mean that there would be layers
and no momentum or energy could go from layer to layer. In
practice this essentially means that one layer cannot communicate
with the next, at least not in a momentum transferring way. That
would presumably mean that the communication between layers would
only be of the form that one layer could influence effective
couplings on neighbouring layers. However no special communication
to particular points in one layer should be possible from the
other layers!

\subsection{What is so special about the 3 space dimensions ?}

A very remarkable prediction from our models is that the number of
spatial dimensions, in which the speed of motion will be
appreciable, is just {\em three}, in wonderful agreement with the
experimental fact that we have just three spatial dimensions. It
should be stressed that we even get this result twice, namely both
for the boson and the fermion cases! Is there a simple reason why
we should get the number three
\cite{MankocBorstnik:2001yz}-\cite{Mirman:1984bx} each time? Well,
we should think of what would have happened if we should have got
a higher dimension, bearing in mind that we have {\em not }
assumed rotational invariance and have effectively used a generic
Hamiltonian. It follows that there should thus be great difficulty
in obtaining degenerate levels because of level repulsion.

In theories with Lorentz (including rotational) invariance there is,
of course, no need for any fine-tuning
to have the different spin (or helicity for massless particles)
states be degenerate for a given momentum. This is because they
have different values for the angular momentum component, which in
such theories is a conserved quantity. But now in our model we do
{\em not} put angular momentum in as an {\em a priori} conserved
quantity. Thus we should not expect that states only distinguished
by their -- for us {\em a priori} non-conserved -- angular
momentum component to be degenerate.

Now the little group for a massless particle -- and we have seen
that the particles predicted in our model are at least at first
massless -- in $(D-1)+1$ dimensions is $SO(D-2)$ extended by the
inclusion of some non-compact Lorentz transformations. The
extensions mentioned are not used for separating spin or helicity
eigenstates and we shall ignore them here. Thus the important part
$SO(D-2)$ of the little group for the massless particle is
non-abelian for all space-time dimensions $D>4$. Only for the spatial
dimension $d=D-1 =3$ or smaller will a non-trivial spin not
automatically imply degeneracy of the different states. The point
is that non-trivial representations of $SO(D-2)$ for $D-1>3$
necessarily have representation dimensions bigger than unity,
which means degeneracy. If rotation is not a good conserved
symmetry, such a degeneracy should never occur. Thus, for $D>4$,
we should not be successful with a model like ours in reproducing
Lorentz invariance, except for scalars which anyway will lack the
mass protection needed to make them accessible to the ``poor
physicist''. Therefore we could not possibly have got more than
the experimental number of space dimensions $D-1 =3$. Actually, as
we discussed in section \ref{cpt},  the CPT-theorem poses a
problem at first even in $D-1 =3$ dimensions: The anti-particle of
say the Weyl particle which we did obtain is in fact degenerate by
CPT symmetry to the original particle. By a similar argument, such
a degeneracy would be expected to be impossible in our model. This
is indeed true but, due to an effective charge conservation, the
two states can be distinguished by an effectively conserved
quantity, i.e.~the charge, and cannot mix. Were it not for this
charge, the particle should have become a very heavy Majorana
particle not accessible to the poor physicist and the degeneracy
would have disappeared also.

Even in the photon case, there must have been an effective charge
for otherwise the level repulsion would have made it impossible
to get the usual CPT-invariant picture out. In this case, it is a
duality transformation generating charge that does the job.

There is a natural mechanism for
these abelian symmetries to arise accidentally, due to particles
carrying a large fundamental scale unit of momentum in addition
to the physical momentum observable by the poor physicist.
However it may be much harder, in such an accidental way, to obtain
non-abelian ``rotation'' symmetries that could prevent the level
repulsion between different components of the little group
representation. Such non-abelian symmetries would be needed
for the case of higher than $ 3 + 1$ dimensions, since then
the rotation group about the direction of motion of a particle
would be non-abelian.
In this way we understand why the number of spatial
dimension predicted by us is at most three.

\section*{Acknowledgements}
 We should like to thank D.~Bennett,
N.~Mankoc Borstnik, A.~Kleppe, D.~Lukman and S.~Rugh for
useful discussions at the
Bled workshops, where we presented an earlier version of this work.
HBN would also like to acknowledge some discussions with
S.~Chadha and M.~Ninomiya on this subject several years ago.

\end{document}